\documentclass[aps,prl,reprint,superscriptaddress,nofootinbib]{revtex4-2}

\usepackage{graphicx}
\usepackage{mathrsfs}
\usepackage{dcolumn}
\usepackage{bm}
\usepackage{amsmath}
\usepackage{upgreek}
\usepackage[colorlinks,linkcolor=red,citecolor=blue,urlcolor=red]{hyperref}
\usepackage[utf8]{inputenc}
\usepackage[T1]{fontenc}
\usepackage{mathptmx}
\usepackage[normalem]{ulem}
\usepackage[colorinlistoftodos]{todonotes}
\usepackage[mathlines]{lineno}
\usepackage{physics}
\usepackage{booktabs}
\usepackage{threeparttable}
\usepackage{array}
\usepackage{longtable}
\usepackage{multirow}
\renewcommand{\t}[1]{\mathrm{#1}}

\hypersetup{
  pdftitle  = {Optomechanical Accelerometer Search for Ultralight Dark Matter},
  pdfauthor = {M. Dey Chowdhury, J. P. Manley, C. A. Condos, A. R. Agrawal, D. J. Wilson}
}

\pdfoutput=1

\begin{document}

\title{Optomechanical Accelerometer Search for Ultralight Dark Matter}

	\author{M. Dey Chowdhury}
     \email{mituldc@arizona.edu}
	\affiliation{Wyant College of Optical Sciences, University of Arizona, Tucson, AZ 85721, USA}

	\author{J. P. Manley}%
    \email{john.p.manley@nist.gov}
	\affiliation{Wyant College of Optical Sciences, University of Arizona, Tucson, AZ 85721, USA}
	\affiliation{National Institute of Standards and Technology, Gaithersburg, MD 20899, USA}

    \author{C. A. Condos}%
	\affiliation{Wyant College of Optical Sciences, University of Arizona, Tucson, AZ 85721, USA}

    \author{A. R. Agrawal}
    
   \author{D. J. Wilson}
	\email{dalziel@arizona.edu}
	\affiliation{Wyant College of Optical Sciences, University of Arizona, Tucson, AZ 85721, USA}
	
	\date{\today}

	\begin{abstract}
  Cavity optomechanical systems have recently been proposed as detectors for ultralight dark matter, leveraging their ability to cool and probe mechanical oscillators at the quantum limit.  Here we present a resonant search for ultralight dark matter using a cavity optomechanical accelerometer. 
    The detector consists of a cryogenic Si$_3$N$_4$-membrane cavity mounted to a 4 K copper plate, with photothermal tuning used to scan its 39 kHz mechanical resonance. Shot-noise-limited displacement readout and radiation-pressure feedback cooling yield an acceleration sensitivity of 10 $\t{n}g_0/\sqrt{\t{Hz}}$ over 30 Hz near resonance.  The detector’s material inhomogeneity gives access to direct vector coupling to the dark-matter field. We conduct a 
    \textcolor{black}{resonant} search based on matched-filter statistics,
 yielding upper bounds consistent with thermal noise and above those set by equivalence principle tests. No signal is observed, but the experiment demonstrates stable, quantum-limited operation and validates a scalable approach to resonant detection. With optimized test masses, lower temperature, and multiplexed arrays, the platform offers a path toward competitive constraints on vector-mediated dark-matter interactions.
    \end{abstract}
	\maketitle

Dark matter is one of the enduring scientific mysteries of our time.  Astrophysical observations suggest that an unidentified particle or class of particles accounts for $\sim\!85\;\%$ of the Universe's gravitating matter content; however, their specific mass and the nature of their coupling to Standard Model (SM) particles remains unknown, despite over 40 years of searching using a variety of direct and indirect techniques \cite{bertone2018new}.

Advances in cooling and probing of solid state mechanical oscillators using optical cavities \cite{aspelmeyer2014cavity} have spurred widespread interest in using such cavity optomechanical systems as tabletop dark matter (DM) detectors \cite{carney2021mechanical,antypas2022new,arvanitaki2016sound,manley2020searching,manley2021searching,hirschel2024superfluid,baker2024optomechanical,deshpande2024demonstration}, leveraging compatibility with cryogenics and rapidly maturing techniques that enable operation at fundamental quantum noise limits.  Particle- and wave-like DM detectors have been proposed, based on impulsive \cite{carney2020proposal} and continuous \cite{carney2021ultralight} force measurement protocols, respectively.  Massive arrays of optomechanical sensors might serve as DM track detectors \cite{carney2020proposal,blanco2022models} or phased antennae arrays \cite{derevianko2018detecting}, and have inspired investigation as an application of entanglement-enhanced distributed force sensing \cite{brady2022entangled,brady2023entanglement}. 

    In this Letter, we describe a search for wavelike, ultralight dark matter (UDM) using a cavity optomechanical system operating as an accelerometer.  The concept of accelerometer-based UDM detection was introduced by \cite{graham2016dark} and later reframed as an optomechanical force sensing problem by~\cite{carney2021ultralight,manley2021searching}.  The basic premise is that DM particles with mass $m_\t{DM}\lesssim 10$ eV/c$^2$ \cite{jackson2023search} 
     (determined by the local DM energy density $\rho_\t{DM}$) coherently combine to create an oscillating field, and that this field may act on SM atoms similar to the electromagnetic force---but with electric charge replaced with a generalized charge such as baryon minus lepton (B-L) number.  As such, \textcolor{black}{within the coherence time,}
     two free-falling SM objects would experience a differential acceleration proportional to $\sqrt{\rho_\t{DM}}$ and their fractional charge difference~$\Delta_{12}$~\cite{manley2021searching,carney2021ultralight,graham2016dark}:
    \begin{equation}\label{eq:aDM}
    a_\t{DM}(t) = g \Delta_\t{12}a_0\cos[\omega_\t{DM}t +\textcolor{black}{\phi_\t{DM}}],
    \end{equation}
where $g$ is an unknown coupling constant, $\omega_\t{DM}$ is the UDM Compton frequency, $a_0 = 2.1\times 10^{11}\;\t{m}/\t{s}^2$ \textcolor{black}{\cite{a0}} is a constant proportional to $\sqrt{\rho_\t{DM}}$, and $\phi_\t{DM}$ is a random phase \textcolor{black}{reflecting the velocity dispersion of DM particles in the Galactic halo.}

\begin{figure}[b!]
		\vspace{-2mm}
		\centering  \includegraphics[width=1\columnwidth]{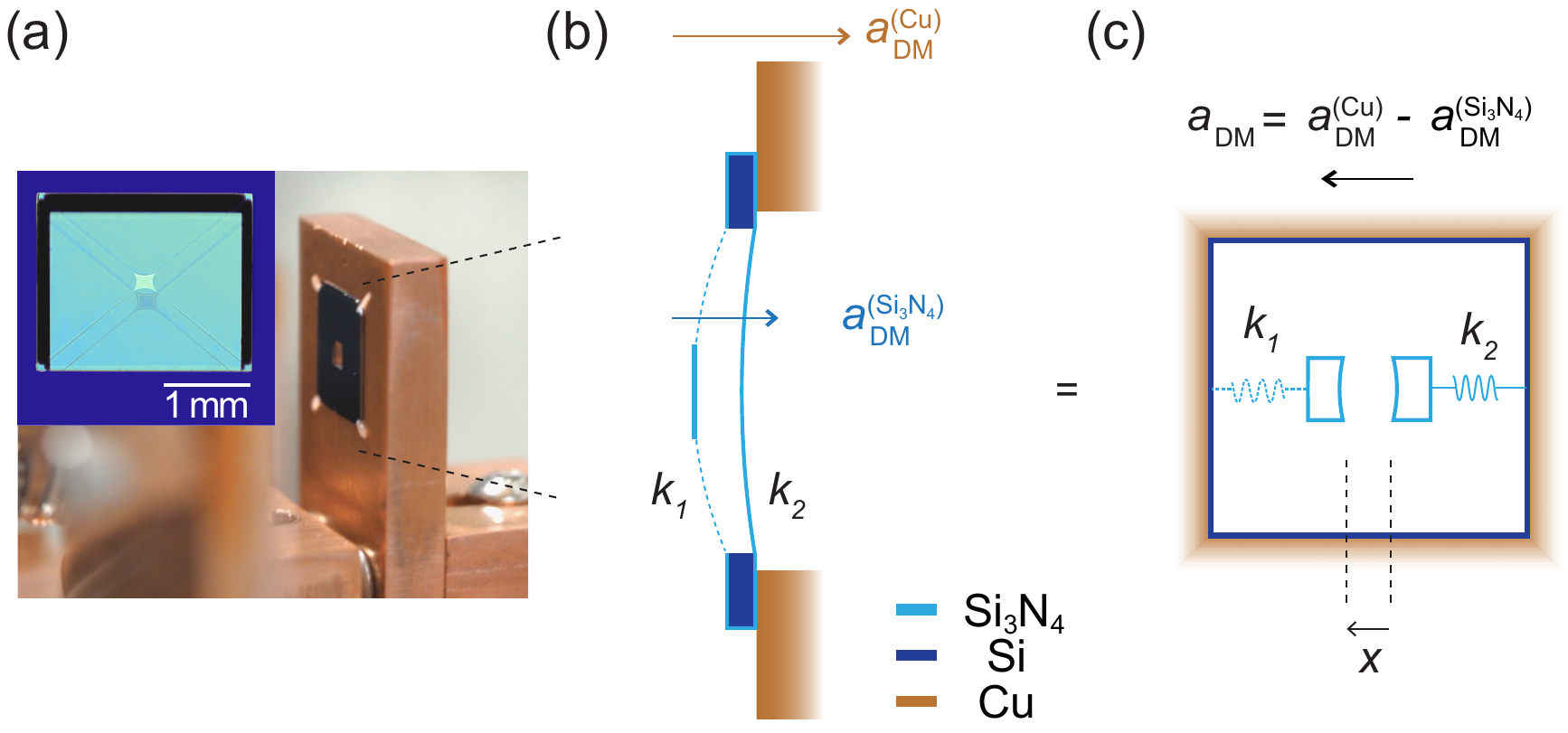}
		\caption{ Dual-membrane optomechanical dark matter detector. (a)~Si chip with suspended Si$_3$N$_4$ membranes fixed to a Cu plate. Inset: Microscope image of trampoline (foreground) and square membrane (background). 
        Concept: (b) UDM differentially accelerates the Cu plate, Si$_3$N$_4$ membranes, and Si chip \cite{SI}. (c) In the reference frame of the base, the membranes experience a base excitation. Their different stiffnesses $k_{1,2}$ result in a relative displacement~$x$~\cite{chowdhury2023membrane}.}
		 \vspace{-4mm}
		\label{fig:1}
\end{figure}

\begin{figure*}[t!]
		\centering  \includegraphics[width=2\columnwidth]{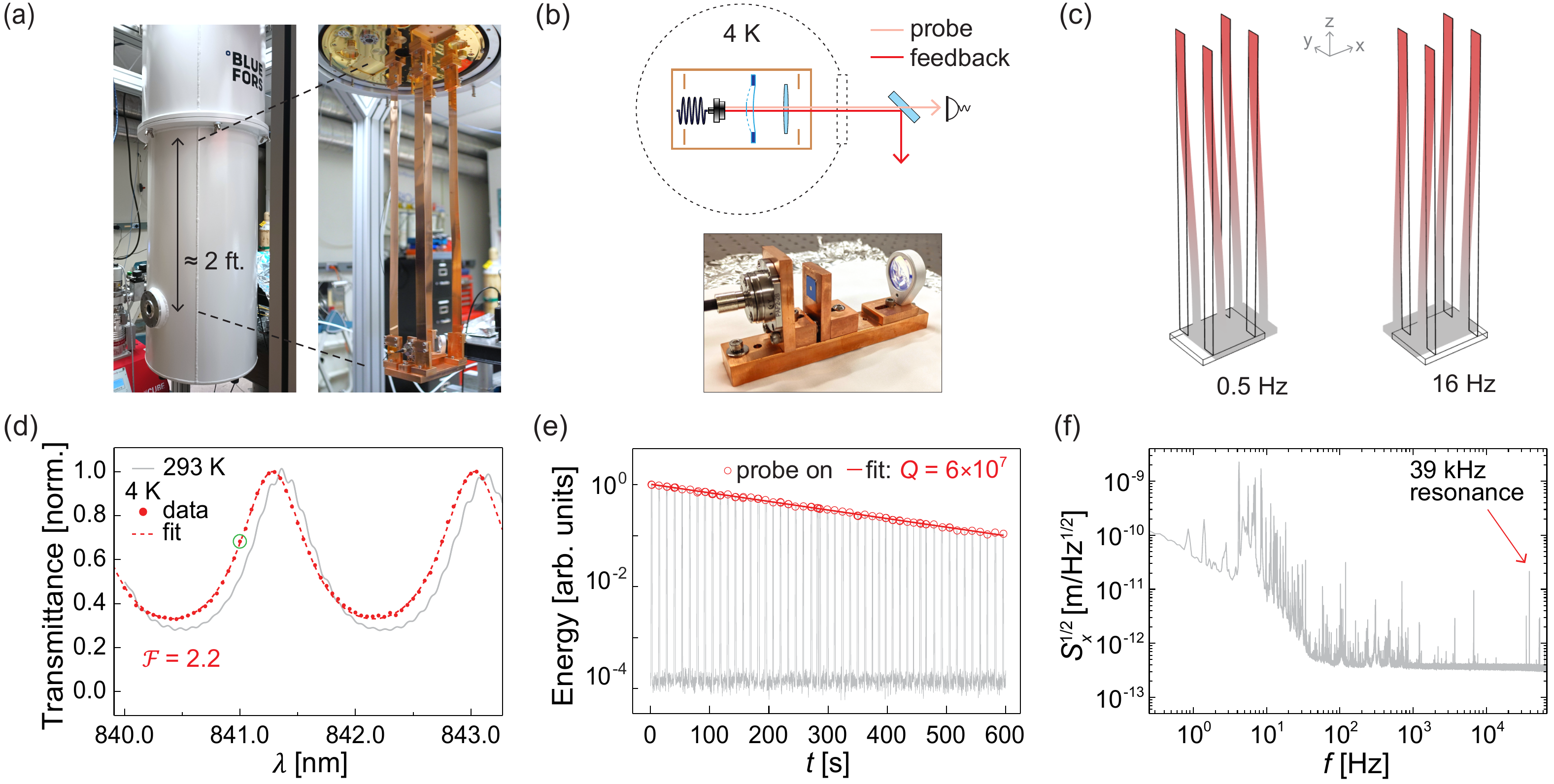}
		\caption{Cryogenic operation enabled by vibration isolation. (a) Photos of Bluefors LD-4K cryostat (left) and custom vibration isolation system (VIS) based on thin Cu strips (right) suspended from the mixing stage. (b) Cavity alignment module (photo, bottom) and readout: probe and feedback laser beams are fiber-coupled and aligned to the dual-membrane cavity; the assembly is mounted on the VIS-platform (top, cartoon). A photodetector placed outside the cryostat records the cavity's transmission. (c) Finite-element simulations of the fundamental modes of the VIS along $x$ (left) and $y$ (right). (d) Transmission fringes obtained from a laser detuning sweep before and after cooldown, showing cavity alignment remains stable. The green circle highlights the detuning used for side-of-fringe readout. (e) Energy ringdown of the trampoline's fundamental mode at 4 K reveals $Q_\t{m}$ of 60 million. (f) Broadband spectrum showing a reduction of vibration background above $f\sim 1\;\t{Hz}$.}\label{fig:2}
\end{figure*}

    The design of our optomechanical accelerometer was proposed in \cite{manley2021searching} and experimentally realized in \cite{chowdhury2023membrane}, and consists of a pair of Si$_3$N$_4$ membranes with different stiffnesses---a trampoline and a square membrane---vertically integrated on a Si chip, forming a Fabry-Perot cavity.  As illustrated in Fig.~\ref{fig:1}, fixing the cavity to a Cu plate (possessing a larger B-L [neutron] density than Si$_3$N$_4$ and Si) translates the UDM signal into an effective chip acceleration  $a_\t{DM}$.  If coincident with the resonance frequency of the trampoline $\omega_\t{m}$, the acceleration gets amplified by the mechanical quality factor into a displacement $x_\t{DM} = Q_\t{m}a_\t{DM}/\omega_\t{m}^{2}$ \textcolor{black}{\cite{xDM}}, yielding sensitivity to UDM with a thermal-noise-equivalent coupling strength of~\cite{manley2021searching}
    \begin{equation}\label{eq:g_DM_th}
        g^\t{(th)} \approx \sqrt{\frac{ 2 k_B T}{m Q_\t{m} Q_\t{DM}}}\frac{\omega_\t{DM}}{\Delta_{12}a_0}\times\left(\frac{2Q_\t{DM}}{\omega_\t{DM}\tau}\right)^{1/4}
    \end{equation}
    where $T$ is the bath temperature, $m$ is the effective mass of the trampoline (accounting for its modeshape and the finite stiffness of the square membrane \cite{manley2021searching}), $Q_\t{DM}\sim 10^6$ \cite{tilde} is the effective $Q$ factor (number of coherent cycles) of the UDM field, and $\tau$ is the measurement time, assuming $\omega_\t{DM}\tau>Q_\t{DM}$.

    Equation \ref{eq:g_DM_th} illustrates the desirability of a high $m\times Q_\t{m}$ mechanical oscillator operated at cryogenic temperatures for sustained intervals.  To realize a novel sensitivity, \cite{manley2021searching} proposed an $\omega_\t{m}\sim 2\pi\times 1$ kHz, $Q_\t{m}\sim 10^9$, cm-scale Si$_3$N$_4$ membrane ($m\sim 1\;\t{mg}$) operating in a closed-cycle dilution refrigerator ($T \sim 10$ mK) for a year ($\tau\sim 10^7$ s), enabling $g_\t{DM}^\t{(th)}\sim 10^{-25}$
    ---two orders of magnitude below current constraints set by precision torsion balance tests of the equivalence principle~\cite{wagner2012torsion,EotWash}.

\begin{figure}[ht!]
		 \vspace{-2mm}
		\centering  \includegraphics[width=1\columnwidth]{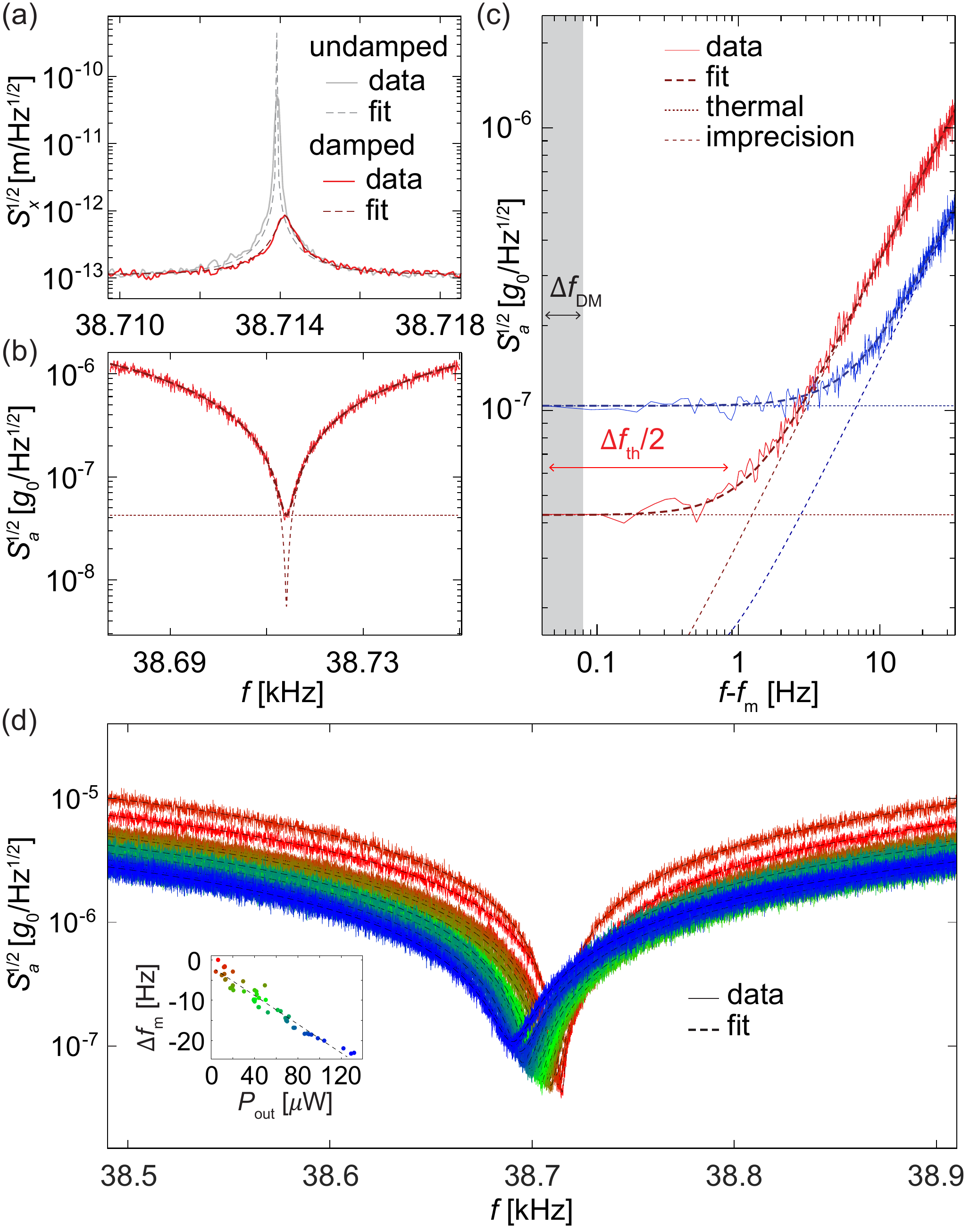}
        \caption{Displacement and acceleration measurements. (a) Calibrated membrane-trampoline displacement PSD near trampoline's  fundamental mode frequency $f_\t{m}$, for a weak probe with transmitted power $P_\t{out} \approx 1\;\upmu\t{W}$.  The red (gray) trace is obtained with (without) optical damping. (b) Corresponding closed-loop acceleration PSD after inverting the mechanical susceptibility. (c) Plots of acceleration noise versus  frequency detuning from resonance at $P_\t{out} = 1\;\upmu\t{W}$ (red) and $100\;\upmu\t{W}$ (blue), illustrating the trade-off between bandwidth and sensitivity due to photothermal heating. The gray-shaded region corresponds to the UDM-signal linewidth $\gamma_\t{DM}\approx 2\pi\times 0.08$ Hz. (d) Photothermal frequency tuning of $f_\t{m}$. Increasing the probe power \textcolor{black}{(inset) decreases $f_\t{m}$ at the expense of increased thermal noise $S_a^\t{th}$ \cite{SI}.}}\label{fig:3}
        
\end{figure}

    Here we report a more modest, first generation search for UDM employing a $\omega_\t{m}\sim 2\pi\times 10$ kHz, $Q_\t{m}\sim 10^8$, millimeter-scale Si$_3$N$_4$ trampoline ($m\sim 10\;\t{ng}$) operating in a closed-cycle 4~K cryostat for $\tau \sim 10^3$ s, targeting $g_\t{DM}^\t{(th)}\sim 10^{-17}$. Our aim is to demonstrate the working principle behind a cavity optomechanical vector UDM detector, highlighting the challenges of vibration isolation, how to improve signal averaging using radiation pressure feedback cooling, strategies to extend bandwidth exploiting quantum-noise-limited readout and photothermal frequency tuning, and routines for data analysis by template matching to a model for the UDM signal.

    An overview of our cryogenic apparatus and device characterization is shown in Fig. \ref{fig:2}. In conceiving such a system, we first emphasize the challenging acceleration sensitivity requirements implied by current UDM constraints in the ``optomechanics'' band, $f_\t{DM} =\omega_\t{DM}/2\pi\sim 1\;\t{Hz}$ to $1\;\t{MHz}$. For B-L UDM, for example, $g\lesssim 10^{-22}$ (see Fig. 4) corresponds to an acceleration sensitivity of $\sqrt{S_a}\sim g a_0\Delta_{12}\sqrt{\omega_\t{DM}/Q_\t{DM}}\lesssim \Delta_{12}\times10^{-11}g_0/\sqrt{\t{Hz}}$, where $g_0=9.8\;\t{m}/\t{s}^2$ is Earth's standard gravity.  This sensitivity has been achieved by the Laser Interferomeric Gravitational Wave Observatory (LIGO) at frequencies below 1 kHz---indeed, B-L constraints from $100\;\t{Hz}$ to $2\;\t{kHz}$ currently belong to LIGO---however, at higher frequencies, we know of no such reported acceleration sensitivity, including with mechanical oscillators.  Challenges include unfavorable thermal noise scaling $S_a^\t{th}=4k_B T\omega_\t{m}/(m Q_\t{m})$, ambient vibrations produced by cryostats, and difficulty thermalizing micromechanical resonators to sufficiently low $T$, while at the same time being able to resolve their  thermal motion.

    We adopt a conservative approach and house our detector in a field-upgradable Bluefors LD-4K cryostat (LD with the dilution unit removed) \cite{NISTdisclaimer}, using the extra sample volume for a two-foot-long (1 foot = 1 ft. = 30.5 cm) pendulum vibration isolation system (VIS) \cite{planz2023membrane} (Fig. \ref{fig:2}a).  The frequency of the pendulum, $f_\t{p} = 0.5$ Hz, is five orders of magnitude below the trampoline's $f_\t{m} = 39$ kHz, implying access to an isolation factor of $(f_\t{m}/f_\t{p})^2 \sim 10^{10}$.  To mitigate pendulum swing, the accelerometer and fiber-optic delivery system are assembled and pre-aligned on a small optical bench mounted on the sample stage (Fig. \ref{fig:2}b).  The cavity transmission is monitored through a viewport by a low-noise Si photodetector. As shown in Fig. \ref{fig:2}d, the cavity alignment survived cooldown with minimal loss of visibility, \textcolor{black}{realizing the design finesse $\mathcal{F}=2.2$~\cite{finesse}.}  A mechanical ringdown revealed $Q = 6\times 10^7$ \textcolor{black}{(Fig. \ref{fig:2}e)} and thermal noise measurements (see below and \cite{SI}) confirmed that \textcolor{black}{the fundamental trampoline mode} thermalized to the cryostat base temperature of 3.5 K, corresponding to $\sqrt{S_a^\t{th}} = 3\times 10^{-8}\;g_0/\sqrt{\t{Hz}}$, using $m = 12\;\t{ng}$~\cite{chowdhury2023membrane}.

    To search for UDM, we analyze acceleration spectra for the presence of a narrow spectral feature with width $\gamma_\t{DM}=\omega_\t{DM}/Q_\t{DM}$.
    Toward this end, Fig. \ref{fig:3} shows near-resonance estimates of the trampoline-membrane displacement power spectral density (PSD) $S_x[\omega]$ and, by inference, chip acceleration PSD  $S_a[\omega]$, using a standard feedback-assisted force sensing protocol \cite{chowdhury2023membrane,gavartin2012hybrid,pontin2014detection,gisler2024enhancing}. For these measurements, the cavity was probed with a low noise wavelength-tunable (840-852 nm) diode laser, for displacement readout, and an auxiliary fixed wavelength (658 nm) diode laser, for radiation pressure feedback. The readout laser was tuned to the side of the fringe (Fig. \ref{fig:2}e) to maximize sensitivity. The feedback laser was intensity modulated with a delayed copy of the readout photocurrent to damp the resonance by $\gamma>\gamma_\t{DM}$~\cite{ColdDamping} (Fig. \ref{fig:3}a). Periodograms were then averaged for $\tau =  800\;\t{s}$ in $N = 40$ intervals of $ \tau_\t{p} = 20\;\t{s}\approx 10\gamma_\t{DM}^{-1}$ \textcolor{black}{\cite{MeasurementTime}}, calibrated in displacement units using the fringe slope \cite{chowdhury2023membrane}, and divided by the closed-loop acceleration susceptibility $\chi[\omega] = (\omega^2-\omega_\t{m}^2+i\gamma\omega_\t{m})^{-1}$, yielding an acceleration PSD estimate $\hat{S}_a[\omega]$ with an ideal (Gaussian-distributed) mean and standard deviation of 
     \begin{subequations}\begin{align}\label{eq:Sa}
     S_a[\omega] &= S_a^\t{DM}[\omega]+S_a^\t{th}+|\chi[\omega]|^{-2} S_x^\t{imp}\;\t{and}\\\label{eq:sigmaSa}
     \sigma_{\hat{S}_a}[\omega] &= S_a[\omega]/\sqrt{N}
    \end{align}\end{subequations}
    respectively, where $S_x^\t{imp}$ is the displacement readout noise (imprecision), and $S_a^\t{DM}[\omega]$ is the hypothetical UDM signal.

    As shown in Fig. \ref{fig:3}, for readout powers from $P_\t{out}\approx 1\;\upmu\t{W}$ to $100\;\upmu\t{W}$, we realize a shot-noise-limited displacement imprecision of $\sqrt{S_x^{\text{im\smash{p}}}}\approx (10^{-13}\text{--}10^{-14})\;\t{m}/\sqrt{\t{Hz}}$, corresponding to a thermal-noise-limited bandwidth of $\Delta \omega_\t{th}=\omega_\t{m}\sqrt{S_x^\t{th}/S_x^{\text{im\smash{p}}}}\approx 2\pi\times 2\;\t{Hz}$, or $\Delta\omega_\t{th}/\gamma_\t{DM}\approx 26$ independent DM bins.  We also observe photothermal heating at the level of $d T/dP_\t{out}\approx 0.4\;\t{K}/\upmu\t{W}$ \textcolor{black}{\cite{photothermalheating}.  We leverage this heating to photothermally tune} the resonance frequency over 25 Hz (330 DM bins), at the expense of a 3-fold increase in \textcolor{black}{thermal noise $\sqrt{S_a^\t{th}}$ \cite{SI}.}

Our search algorithm \cite{SI} involves matched-filtering $\hat{S}_a[\omega]$ to a model $S_a^{\mathrm{DM}}[\omega] = \langle a_\t{DM}^2\rangle G_\t{DM}[\omega]$, where $\langle a_\t{DM}^2\rangle$ is the unknown UDM acceleration power and $G_\t{DM}$ is the normalized lineshape ($\int G_\t{DM}[\omega]d\omega/(2\pi) = 1$). Towards this end, we consider an astrophysically motivated lineshape \cite{derevianko2018detecting,antypas2022new}
\begin{equation}
    G_\t{DM}[\omega] \approx   \frac{\sqrt{8\pi}}{e\gamma_\t{DM}}e^{-\tfrac{2(\omega-\omega_\t{DM})}{\gamma_\t{DM}}}\sinh\sqrt{1+\tfrac{4(\omega-\omega_\t{DM})}{\gamma_\t{DM}}}
\end{equation}
 reflecting the Maxwell-Boltzmann velocity distribution of massive bodies in the Milky Way. We implement a 45-step photothermal frequency scan by increasing the optical power from $5\;\upmu$W to $132\;\upmu$W. We then construct a composite estimator for $\langle a_\t{DM}^2\rangle \equiv D$ based on matched filtering (over frequency index $k$) and weighted averaging (over scan index~$i$)
\begin{equation}\label{eq:D}
    \hat{D}[\omega_\t{DM}] = \sigma^2_{\hat{D}}[\omega_\t{DM}]\sum_{ik} \frac{G_\t{DM}^{\tau_\t{p}}[\omega_k,\omega_\t{DM}]}{\sigma^2_{\hat{S}^{(i)}_a}[\omega_k]}\left(\hat{S}^{\smash{(i)}}_a[\omega_k]-S_a^{\t{n},(i)}[\omega_k]\right)
\end{equation} where  
$G_\t{DM}^{\tau_\t{p}}$ is UDM lineshape binned over $\tau_\t{p}^{-1}$ and $S_a^\t{n} = S_a - S_a^\t{DM}$ is the total measurement noise.  Applying frequentist and Bayesian frameworks \textcolor{black}{\cite{FreqBay2}}, respectively, we implicitly define a detection threshold $D_\t{DT}$ and an upper bound for the DM signal power $D_\t{UL}$, based on a confidence level
\begin{equation}\label{eq:CL}
\t{CL} = \left(\int_{-\infty}^{D_\t{DT}}\rho(\hat{D}|D=0)d \hat{D}\right)^{N_\t{b}} = \int_{-\infty}^{D_\t{UL}}\rho(D|\hat{D})d D,
\end{equation}
where $\rho(\hat{D}|D=0)$ and $\rho(D|\hat{D})$ are the likelihood function and posterior distribution for the DM signal power, respectively, and $N_\t{b}$ is the number of independent frequency bins in the search span, accounting for the look-elsewhere effect~\cite{SI}.

     \begin{figure*}[t!]
		\vspace{-3mm}
		\centering  \includegraphics[width=1.6\columnwidth]{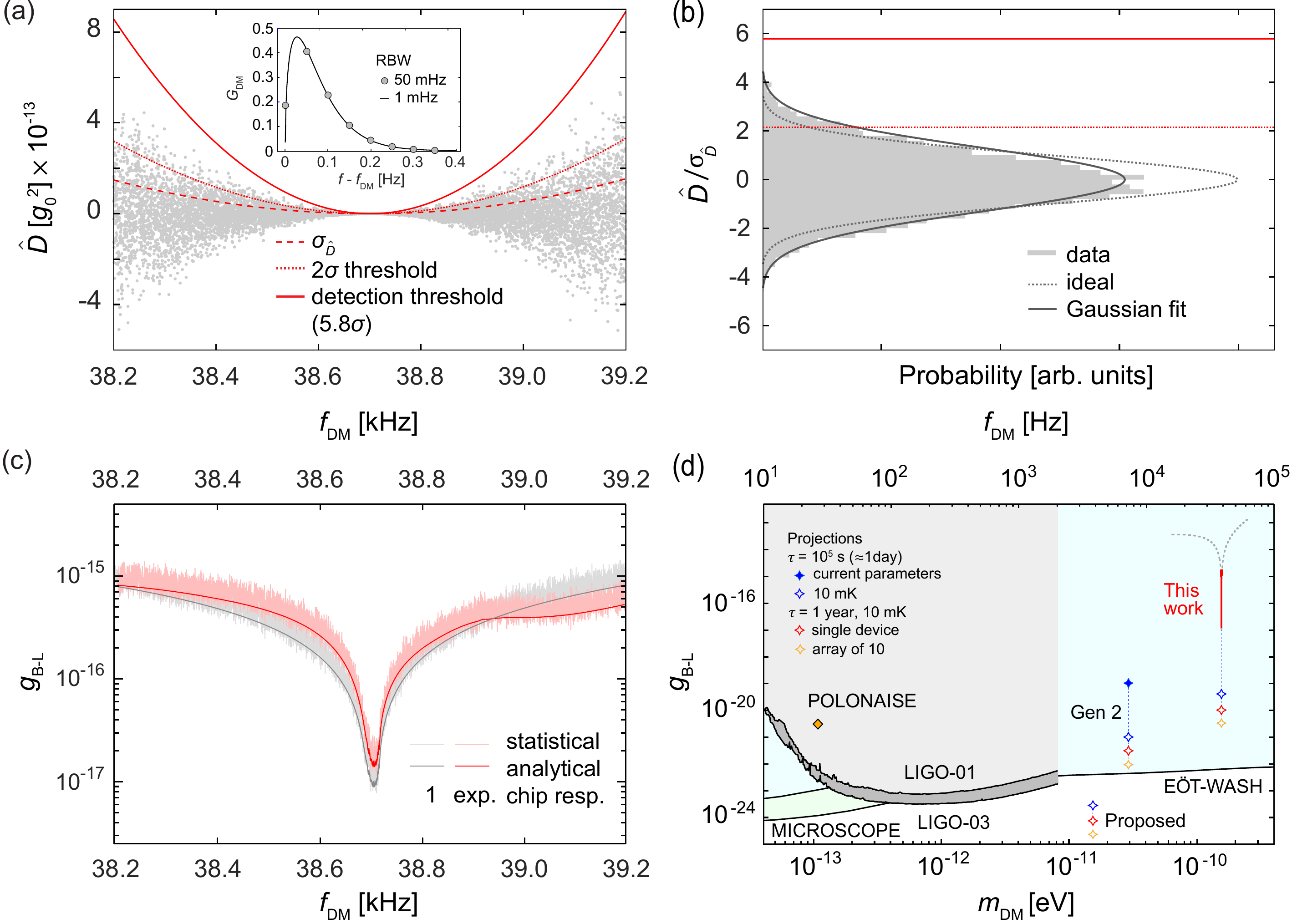}		\caption{Search for B-L UDM. (a) Estimates of UDM acceleration power $\langle a_\t{DM}^2\rangle \equiv D$ obtained from the data in Fig. \ref{fig:3}d using a matched-filter and weighted averaging, Eq.~\ref{eq:D}.  Dashed lines are thresholds for confidence level $\t{CL} = 68\;\%$ ($1\sigma$) and $95\;\%$ ($2\sigma$).  The solid red line is our threshold, which includes the look-elsewhere effect (Eq. \ref{eq:CL}). (b) Histogram of normalized power estimates $\hat{D}/\sigma_D$. Solid and dotted curves show measured and ideal distributions expected from periodogram averaging (Eq. \ref{eq:sigmaSa}). (c) Constraints on $g_\t{B-L}$ from data in (a). Light and bold traces are statistical and analytic estimates, respectively. Gray traces assume a rigid chip; red traces include the estimated chip transfer function~\cite{SI}. (d) Comparison to current constraints from LIGO \cite{guo2019searching,abbott2022constraints}, the E\"{o}t-Wash experiment \cite{wagner2012torsion}, MICROSCOPE \cite{berge2018microscope}, and POLONAISE \cite{amaral2025first}. Gray dashed line is an extension of the analytical model in (c) assuming a displacement imprecision of $S_x^\t{imp}=(5\times 10^{-14}\;\t{m}/\sqrt{\t{Hz}})^2$ \cite{SI}. \textcolor{black}{Markers indicate benchmark sensitivities for second-generation \cite{gen2} and long-term proposed \cite{manley2021searching}  configurations discussed in the main text and \cite{SI}.}} 
		\label{fig:4}
        \vspace{-2mm}
	\end{figure*}

Results of the UDM search are summarized in Fig. \ref{fig:4}.  As shown in Fig. \ref{fig:4}(a,b), we observe a near-Gaussian distribution of weighted estimators $\hat{D}[\omega_k]/\sigma_{\hat{D}}[\omega_k]$, with variance $\sigma^2_{\hat{D}}[\omega_k]$ \cite{SI} minimized over the $ 25$ Hz resonant scan window in Fig. \ref{fig:3}d. This motivates a likelihood function $\rho(\hat{D}|D)\propto \t{exp}[(D-\hat{D})^2/2\sigma^2_{\hat{D}}]$, from which we determine the 
$\t{CL} = 95\;\%$ confidence threshold shown in Fig.~\ref{fig:4}a.  Accounting for the look-elsewhere effect over a search span of $\Delta f = 1\;\t{kHz}$ ($N_\t{b} = \tau_\t{p}\Delta f  = 10^4$), \mbox{we observe no detection events.}

To set a limit on the DM coupling strength, we combine the experimentally determined likelihood function with Bayes theorem and a prior $\rho(D)$ that is uniform for $D>0$, 
yielding a median limit $D_\t{UL}\approx \sqrt{2}\sigma_{\hat{D}}\t{Erf}^{-1}[\t{CL}]$ \cite{SI} and a $2\sigma$ bound
\begin{equation}
g\le \frac{\sqrt{2D_\t{uL}}}{a_0 \Delta_{12}}\approx \frac{2\sqrt{\sigma_{\hat{D}}}}{a_0 \Delta_{12}}
\end{equation}
for $\t{CL} = 95\;\%$. We consider B-L coupling ($g_\t{B-L}$) \textcolor{black}{as a canonical example} \cite{fileviez2010baryon,bauer2018hunting} 
in which case $\Delta_{12} = Z_1/A_1-Z_2/A_2$, where $Z_{i}$ and $A_{i}$ are the atomic and mass number of body $i$, respectively.  For Si$_3$N$_4$ and Si, $Z/A \approx 0.50$ and for Cu, $Z/A\approx 0.46$, yielding $\Delta_{12} \approx 0.04$.  Using this value presumes that the accelerometer behaves like a heterogeneous mechanical dimer (two bodies made of different material, attached by a simple spring) \cite{manley2021searching}, as substantiated in  \cite{SI}.

Inferred $g_\t{B-L}$ constraints are shown in Fig. \ref{fig:4}c, assuming a perfectly rigid chip (gray) and correcting for the estimated chip response function (red) \textcolor{black}{(see \cite{SI} for details)}. The lower bound $g_\t{B-L}\approx 1\times 10^{-17}$ corresponds to an acceleration amplitude resolution of $\sqrt{\sigma_{\hat{D}}} \approx 4\times10^{-9}\;g_0$ and PSD resolution of $\sigma_{\hat{S}_a}(\omega_\t{DM}) \approx (2\times 10^{-8}g_0/\sqrt{\t{Hz}})^2$, consistent with the variance of the lowest PSD estimate in Fig. \ref{fig:3} \cite{SI} (at $T\approx$ 8 K due to photothermal heating).  Shown in Fig. \ref{fig:4}d is our bound combined with the best current constraints from the E\"{o}t-Wash experiment \cite{wagner2012torsion,EotWash} and various others at lower frequency. 

\color{black}
Looking forward, realizing the original design sensitivity $g_\t{B-L}\sim10^{-25}$ \textcolor{black}{proposed in} \cite{manley2021searching} requires operating at 10 mK, increasing the cavity finesse to $\mathcal{F}=100$, increasing the size and $Q$ of the membrane to $10$ cm and $10^9$, respectively, vibration isolation at the $10^{-12}\;g_0/\sqrt{\t{Hz}}$ level, and integrating for a year \cite{Sidereal}.
Recent advances augur well for these upgrades:  Sub-100-mK, $\mathcal{F}\gtrsim 10^4$ cavity optomechanics experiments with Si$_3$N$_4$ membranes have been demonstrated~\cite{planz2023membrane,brubaker2022optomechanical}; as have 
high reflectivity photonic crystal~(PtC) membranes \cite{zhou2024ultrahigh}, 
including stable $\mathcal{F}\approx10^3$ cavities employing a gradient-pitch PtC \cite{agrawal2024focusing}. 
Decimeter-scale membranes are now commercially available \cite{Norcada}, and centimeter-scale PtC membranes \cite{moura2018centimeter,norder2025pentagonal} have been realized, driven by the Starshot project.  The expected mechanical $Q$ versus size scaling (due to dissipation dilution \cite{engelsen2024ultrahigh}) from $10^8$ to $10^{10}$ has also been confirmed with centimeter-scale Si$_3$N$_4$ beams \cite{cupertino2024centimeter}. Finally, low vibration, closed-cycle (continuous operation) dilution refrigerator technology continues to advance in response to the superconducting quantum computing industry, with custom $10^{-11} g_0/\sqrt{\t{Hz}}$ VIS demonstrated for frequencies as low as $10$ Hz~\cite{amaral2025first}. \color{black}

A key component to realizing novel contraints will be optimizing the membrane geometry. In-plane patterning can enable ultrahigh-$Q$ \cite{engelsen2024ultrahigh}, however, a promising alternative is 3D-mass-loading \cite{pratt2023nanoscale,zhou2021broadband,bawden2025precision}, which can enhance the more relevant $Q$-$m$ product (Eq. \ref{eq:g_DM_th}).  Towards this end, we highlight a recent demonstration of $m\sim 0.1 \;\t{mg}$, $f_\t{m}\sim 100$ Hz, $Q_\t{m}\sim 10^7$ torsion micropendula based on mass-loaded Si$_3$N$_4$ nanoribbons, which achieve $S_a^\t{th}\sim 10^{-9}\;g_0/\sqrt{\t{Hz}}$ at room temperature~\cite{condos2025ultralow}. Cryogenically cooled, these devices could reach the desired sensitivity for novel UDM searches in the 10 Hz to 1 kHz band, in conjunction with quantum-limited optical lever measurements \cite{pluchar2025quantum,shin2025active}; they also provide a route to vibration isolation through heterogeneous mass-loading~\cite{sun2025differential}, the latter of which has been demonstrated using a flip chip approach \cite{bawden2025precision}.


Finally, we highlight a recent proposal to search for UDM with an array of $N = 10$ cryogenic membrane accelerometers probed by a distributed squeezed light source \cite{brady2023entanglement}, building on an $N = 2$ demonstration with two $0.1$-mm membranes~\cite{xia2023entanglement}. This approach enables enhanced sensitivity (as much as $N$-fold \cite{brady2023entanglement}) if the power per sensor is fixed and, therefore, improved UDM searches with a cryogenic detector limited by photothermal heating or other power constraints. Loss and scalability are key challenges. \textcolor{black}{Recent demonstrations of vertically integrated $\mathcal{F}\approx10^3$ PtC-membrane cavities \cite{khokhar2026high} suggest a path toward scalable arrays using free-space squeezed light}, while emerging photonic-integrated accelerometers \cite{ge2025towards} and squeezed-light sources \cite{clark2025integrated} offer a fully integrated approach.


In the \textcolor{black}{Supplementary Material}, we compare various contemporary optomechanical accelerometers and their projected performance as B-L UDM detectors, including a second-generation, 1~$\upmu$g Si$_3$N$_4$ membrane accelerometer \textcolor{black}{[``Gen-2'' in Fig. \ref{fig:4}d]}, and an emerging class of electromechanical accelerometers based on magnetically levitated test masses, exemplified by the POLONAISE UDM detector \cite{amaral2025first}.  We find that several of these systems could in principle reach competitive sensitivity to vector-mediated DM interactions, when operated for extended periods at millikelvin~temperatures. 
\color{black}

\vspace{-2mm}
\section{Acknowledgments}
\vspace{-2mm}

The authors thank Swati Singh and Daniel Grin for laying the theoretical foundations of this work, Atkin Hyatt for designing and fabricating next generation test masses presented in the Supporting Materials, Christian Pluchar for guidance with laser intensity noise reduction, Andrew Land for help constructing the vibration isolation system, and Utkal Pandurangi for helpful discussions. This work was supported by NSF Grants 2209473 and 1945832 and from the Northwestern University Center for Fundamental Physics and the John Templeton Foundation through a Fundamental Physics grant. 

\vspace{-2mm}
\section{Data Availability Statement}
The data that support the findings of this study are publicly available \cite{ZenodoData}.\vspace{-2mm}


    \bibliography{ref}

\begin{thebibliography}{43}
\bibitem [{NIS()}]{NISTdisclaimer}%
  \BibitemOpen
  \href@noop {} {}\bibinfo {note} {Commercial equipment, instruments, and materials are identified in this paper in order to specify the experimental procedure adequately. Such identification is not intended to imply recommendation or endorsement by the National Institute of Standards and Technology or the United States government, nor is it intended to imply that the materials or equipment identified are necessarily the best available for the purpose.}\BibitemShut {Stop}%
\end{thebibliography}
\begin{thebibliography}{95}
\bibitem [{xDM()}]{xDM}%
  \BibitemOpen
  \href@noop {} {}\bibinfo {note} {\textcolor{black}{This formula is generally valid in the frequency domain, where $x_\t{DM}$ and $a_\t{DM}$ are interpreted as Fourier amplitudes at $\omega_m$. In the time domain, it is valid only if $Q_\t{DM} \gg Q_\t{m}$.}}\BibitemShut {Stop}%
\bibitem [{til()}]{tilde}%
  \BibitemOpen
  \href@noop {} {}\bibinfo {note} {Here and throughout, we use `$\sim$` to mean 'roughly' or 'on the order of.'}\BibitemShut {NoStop}%
\bibitem [{Mea()}]{MeasurementTime}%
  \BibitemOpen
  \href@noop {} {}\bibinfo {note} {\textcolor{black}{Measurement time $\tau=800$ s was limited by practical considerations such as the open stability of our laser and the manual procedure for frequency scanning. Future, longer searches will require more stringent stabilization and automation.}}\BibitemShut {Stop}%
\end{thebibliography}

\end{document}


\title{Supplementary Information for \\Optomechanical Accelerometer Search for Ultralight Dark Matter}
	
	\author{M. Dey Chowdhury}
	\affiliation{Wyant College of Optical Sciences, University of Arizona, Tucson, AZ 85721, USA}

	
	\author{J. P. Manley}%
	\affiliation{Wyant College of Optical Sciences, University of Arizona, Tucson, AZ 85721, USA}
	\affiliation{National Institute of Standards and Technology, Gaithersburg, MD 20899, USA}


    \author{C. A. Condos}%
	\affiliation{Wyant College of Optical Sciences, University of Arizona, Tucson, AZ 85721, USA}

    \author{A. R. Agrawal}
    
   	\author{D. J. Wilson}
	\affiliation{Wyant College of Optical Sciences, University of Arizona, Tucson, AZ 85721, USA}
	
	\date{\today}

    \begin{abstract}
        Below we provide supplementary information about the spectral signature of ultralight dark matter (UDM), the procedure for searching for this signal using a frequency-scanned resonant detector, and the experiment. 
    \end{abstract}
	
	\maketitle

\tableofcontents

\vspace{-4mm}
\section{Vector UDM Signal}
\vspace{-1mm}
We consider a vector DM field $A^\nu$ coupled to a conserved particle flux density $J^\nu$, with a Lagrangian density
\begin{equation}	\label{massiveprocalagrangian}
\mathcal{L}=-\frac{c^2}{4}F_{\mu\nu}F^{\mu\nu}+\frac{{\omega_\t{DM}}^2 }{2} A^\nu A_\nu-g \frac{e}{\sqrt{\epsilon_0}} J^\nu A_\nu.
\end{equation}
where $F^{\mu\nu}$ is the field tensor, and $\epsilon_0$ is the vacuum permittivity, $e$ is the electron charge, and $c$ is the speed of light. 

\vspace{-2mm}
\subsection{UDM Field}
\vspace{-2mm}

Temporal and spectral properties of $A_\nu$ are determined by the general properties of UDM. For simplicity, we consider only \textcolor{black}{the projection of the field onto the preferred axis of the detector and simplify the notation as $A_\nu\rightarrow A$}.  Following \cite{derevianko2018detecting}, we model $A(t)$ as a partially coherent, monochromatic wave\footnote{\textcolor{black}{Resolving the spectral lineshape (Eq. \ref{eq:A_PSD}) requires breaking this approximation, but analyzing the signal power $\langle A^2\rangle$ remains valid in both cases.}} 
\begin{equation}\label{eq:A(t)}
A(t) \approx A_0 \cos\left(\omega_\t{DM}t+\color{black}\phi_\t{DM}\color{black}\right)    
\end{equation}
\textcolor{black}{where $\omega_\t{DM}$ is the Compton frequency. For measurement times much smaller than the coherence time, $\sim\gamma_\t{DM}^{-1}$, $A_0$ ($\phi_\text{DM}$) can be treated as a random variable sampled from a Rayleigh (uniform) distribution~\cite{centers2021stochastic}.}

In the frequency domain, the single-sided power spectral density $S_A$  can be expressed as  
\begin{equation}\label{eq:A_PSD}
S_A[\omega] = \langle A ^2 \rangle G_\t{DM}[\omega]
\end{equation}
where $\langle A^2 \rangle$ is the field power, $G_\t{DM}[\omega]$ is a normalized lineshape ($\int_{0}^\infty G_\t{DM}(\omega)\tfrac{d\omega}{2\pi}= 1$) of the form
\begin{subequations}\begin{align}
       \!\! G_\t{DM}[\omega]&= \frac{4}{\gamma_\t{DM}}\frac{\sqrt{2\pi}}{e}  e^{-\frac{2(\omega-\omega'_\t{DM})}{\gamma_\t{DM}}}\sinh\sqrt{1+\tfrac{4(\omega-\omega'_\t{DM})}{\gamma_\t{DM}}}\\
        & = \frac{4}{\gamma_\t{DM}}\sqrt{\frac{2\pi}{e}}  e^{-\frac{2(\omega-\omega_\t{DM})}{\gamma_\t{DM}}}\sinh\sqrt{\tfrac{4(\omega-\omega_\t{DM})}{\gamma_\t{DM}}}
\end{align}\end{subequations}
for $\omega\geq \omega_\t{DM}$, and $G_\t{DM}[\omega]=0$ for $\omega<\omega_\t{DM}$, and
\begin{equation}
    \begin{aligned}
        \omega'_\t{DM}&\equiv \omega_\t{DM}\left(1+\frac{\xi_\t{DM}^2}{2}\right),\;\t{and} \\
        \gamma_\t{DM}&\equiv 2 \xi_\t{DM}^2 \omega_\t{DM}
    \end{aligned}
\end{equation}
are the Doppler-shifted Compton frequency and spectral width (approximately full width at half maximum) of the UDM wave, respectively, expressed in terms of its velocity dispersion relative to the speed of light, $\xi_\t{DM}\sim 10^{-3}$ \cite{derevianko2018detecting}.

\begin{figure*}[ht!]
    \centering
    \includegraphics[width=2\columnwidth,trim= 0in 0in 0in 0in]{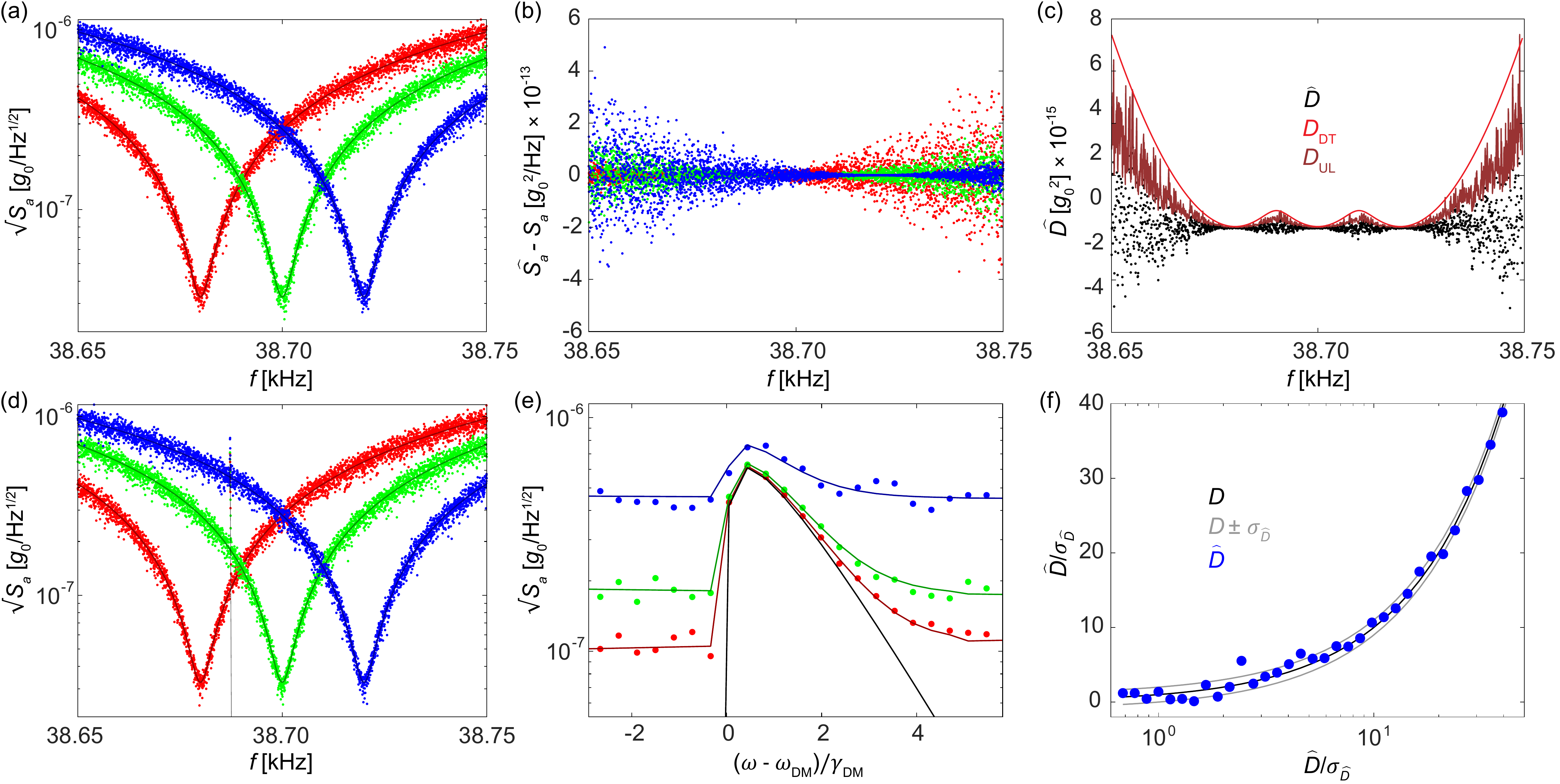}
    \caption{Simulated UDM Search. Top: Drawing constraints from simulated acceleration data containing no UDM signal.  Readout and thermal noise are modeled after cold-damped, frequency-swept device in the main text, Figs. 3-4. (a) Acceleration PSD estimates $\hat{S}_a$ for three resonance frequencies, individually fitted to infer a mean value $S_a$; (b) corresponding estimates of the excess signal power $\hat{S}_a-S_a$; (c) composite estimator $D$ (Eq. \ref{eq:D}), detection threshold $D_\t{DT}$ (Eq. \ref{eq:threshold}), and inferred upper bounds $D_\t{UL}$ (Eq. \ref{eq:DUL}). Bottom: Detecting an injected signal. (d) A persistent UDM signal (Eq.~\ref{eq:aa_PSD}) is injected into the three datasets from (a). (e) Zoom in on the injected signal. (f) Matched filtering estimates are performed for various simulated signal strengths, demonstrating consistency of the estimation procedure.}
    \label{fig:simulatedSearch}
\end{figure*}

The 00-component of the symmetrized stress-energy tensor of the field can be shown to be~\cite{manley2022searching}
\begin{equation}
    T^{00}\approx {\omega_\t{DM}}^2 A^i A_i
\end{equation}
where indices indicate summation over the spatial components of the field. By equating the estimated local UDM energy density $\rho_\t{DM}\approx0.4$ GeV/cm\textsuperscript{3}~\cite{catena2010novel} \textcolor{black}{with the time-averaged tensor component $\langle T^{00}\rangle$, and assuming the field is isotropic $\langle A^2 \rangle = \tfrac{1}{3}\langle A^i A_i \rangle$  \cite{graham2016dark,manley2021searching,amaral2024vector},} it follows that\footnote{\textcolor{black}{Note that our normalization of $\langle A^2 \rangle$ is equivalent to that of \cite{amaral2024vector} when understood as the time-averaged variance of a field component.  In that work, $A_i(t) \approx  \sqrt{\scriptstyle{2\rho_\t{DM}/(3\omega_\t{DM}^2})}\;\alpha_i\cos\left(\omega_\t{DM} t +\phi_{i,\t{DM}}\right)$, $\alpha_i$ is a random variable with $\langle \alpha_i^2\rangle =1$ (Rayleigh-distributed), and the factor of $3$ is from equipartition.}}
\begin{equation}\label{eq:Apower}
   \langle A^2\rangle\approx\frac{\rho_\t{DM}}{{3\omega_\t{DM}}^2}.
\end{equation}

\subsection{Differential Acceleration Signal}
Vector UDM with interactions described by Eq. \ref{massiveprocalagrangian}  will exert a force on particles carrying dark charge---analogous to the electromagnetic Lorentz force. In the non-relativistic limit, it can be shown that the (dark) magnetic contribution to this force is negligible relative to the electric contribution~\cite{manley2022searching}, such that the force exerted on objects with total dark charge $q$\footnote{For example, for B-L coupling, $q$ is the neutron number.} is
\begin{equation}
    F_\t{DM}=-qg \frac{e}{\sqrt{\epsilon_0}} \frac{d A}{dt}\approx gq F_0\sin\left(\omega_\t{DM}t+\color{black}\phi_\t{DM}\color{black}\right)
\end{equation} 
where $F_0 = \sqrt{2\rho_\t{DM}e^2/(3\epsilon_0)} =3.5\times 10^{-16}$ N.

If $q$ is proportional to mass, it is natural to consider the differential acceleration between two objects of mass $m_i$ \cite{manley2022searching}
\begin{subequations}\begin{align}
    a_\t{DM} &= -\left(\frac{q_2}{m_2}-\frac{q_1}{m_1}\right) g \frac{e}{\sqrt{\epsilon_0}} \frac{d A}{dt}\\
    &\approx g\Delta_{12}a_0  \sin\left(\omega_\t{DM}t+\color{black}\phi_\t{DM}\color{black}\right)
\end{align}\end{subequations}
where $a_0 = F_0/m_\t{n}=2.1\times10^{11}\;\t{m}/\t{s}^2$ and
\begin{equation}
    \Delta_{12}= \frac{q_2}{m_2/m_\t{n}}-\frac{q_1}{m_1/m_\t{n}}
\end{equation}
is the difference in the charge per nucleon ratio of each object.

Various coupling channels are possible for vector UDM~\cite{graham2014parametrically,graham2016dark}. For coupling to Baryon minus Lepton number ($B-L$),
\begin{equation}
\Delta_{12}\approx \frac{A_2-Z_2}{A_2}-\frac{A_1-Z_1}{A_1} = \frac{Z_1}{A_1}-\frac{Z_2}{A_2},
\end{equation}
where $Z_i$ and $A_i$ are the average neutron and mass number of each material, respectively.

The associated acceleration PSD (Eq. 4 in the main text)
\begin{equation}\label{eq:aa_PSD}\
        S_a^{\t{DM}}[\omega] = \langle a_\t{DM}^2\rangle G_\t{DM}[\omega]
\end{equation}
is given by replacing $\langle A^2\rangle$ in Eq. \ref{eq:A_PSD} with
\begin{equation}\label{eq:aDMpower}
\langle a_\t{DM}^2\rangle = \frac{g^2 \Delta_{12}^2 a_0^2}{2} = \frac{g^2 \Delta_{12}^2 \rho_\t{DM} e^2}{3 \epsilon_0 m_\t{n}^2}
\end{equation}

\vspace{-1mm}
\section{UDM Search Data Analysis}

A simulated UDM search is shown in Fig. \ref{fig:simulatedSearch}, mirroring the experimental search presented in Figs 3-4 of the main text.  As shown in Fig. \ref{fig:simulatedSearch}a and 3c, search data consists of estimates of the differential acceleration $\hat{S}^{(i)}_a[\omega_k]$ at different Fourier frequencies $\omega_k$ and detector resonance frequencies $\omega_{0,i}$, the latter varied as part of frequency scan.  Each estimate is obtained using Bartlett's method---i.e., averaging $N_i$ consecutive, non-overlapping periodograms of duration $\tau_\t{p}$, corresponding to a resolution bandwidth of $\omega_k - \omega_{k-1} = 2\pi/\tau_\t{p}$. In the absence of a DM signal, the mean and standard deviation of each estimate is ideally $\langle \hat{S}^{(i)}_a(\omega_k)\rangle = S_a^{(i)}(\omega_k)$ and 
$\smash{\sigma_{\hat{S}_a^{(i)}}(\omega_k) = S^{(i)}_a(\omega_k)/\sqrt{N_i}}$, respectively, where 
$S_a^{(i)}(\omega)$
is the true PSD stemming from a combination of thermomechanical and shot noise.  In practice, we find that $\sigma_{\hat{S}_a^{(i)}}$ is $\approx 30\;\%$ larger than ideal, by comparing each $\hat{S}^{(i)}_a(\omega_k)$ to a model for $S^{(i)}_a(\omega_k)$ obtained from a Lorentzian fit (main text Fig. 4b).  We use these experimentally determined values in our DM search, as discussed below.

\subsection{DM signal power estimation via matched filtering}

To search for DM at frequency $\omega_\t{DM}$, we construct a composite estimator for the signal power
\begin{equation}
    \langle a_\t{DM}^2\rangle \equiv D
\end{equation} that combines matched filtering (over frequency index $k$) and weighted averaging (over measurement index $i$):
\begin{equation}\label{eq:D}
    \hat{D}[\omega_\t{DM}] = \sigma^2_{\hat{D}}[\omega_\t{DM}]\sum_{ik} \frac{G_\t{DM}^\t{\tau_\t{p}}[\omega_k,\omega_\t{DM}]}{\sigma^2_{\hat{S}^{(i)}_a}[\omega_k]}\left(\hat{S}^{\smash{(i)}}_a[\omega_k]-S_a^{(i)}[\omega_k]\right)
\end{equation}
where here
\begin{equation}
    G_\t{DM}^\t{\tau_\t{p}}[\omega_k,\omega_\t{DM}]\equiv \tau_\t{p}\int_{\omega_k - \pi/\tau_\t{p}}^{\omega_k+\pi/\tau_\t{p}}  G_\t{DM}[\omega,\omega_\t{DM}]\tfrac{d\omega}{2\pi}
\end{equation}
is the discretized lineshape of the UDM signal (normalized such that $\sum_k G_\t{DM}^\t{\tau_\t{p}}[\omega_k]=\tau_\t{p}$) and 
\begin{equation} \label{eq:DMestimatorVariance}
    \sigma^2_{\hat{D}}(\omega_{\t{DM}}) = \left( \sum_{ik} \left( \frac{G_\t{DM}^\t{\tau_\t{p}}[\omega_k,\omega_\t{DM}]}{\sigma_{\hat{S}_a^{(i)}}(\omega_k)} \right)^2 \right)^{-1}
\end{equation}
is the estimator variance.

For a sufficiently large data set, the likelihood function of the DM signal estimator can be approximated as a Gaussian:\footnote{\color{black} For a large number of averages, stochastic properties of the dark matter field average out, such that the signal power $D$ can be treated as deterministic~\cite{centers2021stochastic,amaral2024vector}.}
\begin{equation} \label{eq:DDMgaussLikelihood}
    \rho (\hat{D} | D) \approx \frac{e^{-(\hat{D} - D)^2/(2\sigma^2_{\hat{D}})}}{\sqrt{2\pi \sigma^2_{\hat{D}}}}, 
\end{equation}
We confirmed this experimentally as seen in main text Fig.~4.

\subsection{Search for statistically significant signals}
We adopt a frequentist approach to define a threshold signal power $D_\t{DT}$ for DM detection at a confidence level CL.  Accounting for the look-elsewhere effect \cite{gross2010trial,rover2011bayesian,baxter2021recommended} over a search bandwidth of $\Delta \omega$, the threshold is implicitly defined as
\begin{equation}
    \t{CL} = \left(\int_{-\infty}^{D_\t{DT}}\rho (\hat{D} | D = 0) d\hat{D}\right)^{N_\t{b}}
\end{equation}
where $N_\t{b} = \Delta\omega/\gamma_\t{DM}$ is the number of independent search bins and $\rho (\hat{D} | D = 0)$ is the likelihood function under the null hypothesis ($D = 0$).

Assuming a Gaussian likelihood function (Eq. \ref{eq:DDMgaussLikelihood}) yields
\begin{equation} \label{eq:threshold}
    D_\t{DT}(\omega_\t{DM}) = \sqrt{2}  \text{Erf}^{-1} \{ 2\text{CL}^{1/N_\t{b}} - 1 \} \sigma_{\hat{D}}(\omega_\t{DM}).
\end{equation}

\subsection{Upper limits on UDM coupling strength}
We apply a Bayesian framework to set an upper limit $D_\t{UL}$ on the UDM signal power $D$ from the posterior distribution $\rho\left(D|\hat{D}\right)$, defined by the condition\footnote{Following \cite{lyons2010comments}, we do not apply the look-elsewhere effect to our limits.}
\begin{equation}\label{eq:DCL}
    \text{CL} = \int_{-\infty}^{D_\t{UL}} \rho\left(D|\hat{D}\right)dD.
\end{equation}\color{black}
The posterior distribution is defined by Bayes' theorem
\begin{equation}
    \rho\left(D|\hat{D}\right)= \frac{\rho (\hat{D} | D)\rho(D)}{\rho\left(\hat{D}\right)}
\end{equation}
where 
\begin{equation}
    \rho\left(\hat{D}\right)\equiv \int_{-\infty}^{\infty}  \rho (\hat{D} | D)\rho(D)dD
\end{equation}
is a normalization factor. Noting that the signal power $D$ must be positive, we adopt a uniform prior over positive values
\begin{equation}
    \rho(D) = 
    \begin{cases} 
        0 & \text{if } D< 0 \\
        \text{constant} & \text{if } D\geq 0 
    \end{cases}
\end{equation}
This corresponds to Jeffreys prior for a Gaussian likelihood (Eq. \ref{eq:DDMgaussLikelihood}) for $D\geq 0$. The resulting posterior distribution is
\begin{equation}
    \rho\left(D|\hat{D}\right) = 
    \begin{cases} 
        0 & \text{if } D< 0 \\
        \frac{2\rho (\hat{D} | D)}{1+\text{Erf}\left[\hat{D}/\sqrt{2\sigma_{\hat{D}}^2}\right]} & \text{if } D\geq 0 
    \end{cases}
\end{equation}

The upper limit $D_\t{UL}$ is obtained by inverting Eq. \ref{eq:DCL}:
\begin{equation}\label{eq:DUL}
    \begin{aligned}
        D_\t{UL} = &\hat{D}+ \sqrt{2} \sigma_{\hat{D}} \times \\
        &\text{Erf}^{-1}\left[1+\left(\text{CL}-1\right)\left(1+\text{Erf}\left[\frac{\hat{D}}{\sqrt{2}\sigma_{\hat{D}}}\right]\right)\right].
    \end{aligned}
\end{equation}\color{black}
Using Eq. \ref{eq:aDMpower}, we then define the coupling strength bound at a confidence level CL as
\begin{equation}\label{eq:gCL}
    g=\frac{\sqrt{2 D_\t{UL}}}{\Delta _{12} a_0} 
\end{equation}

\subsubsection{Analytical approximation for upper limits}
An analytical approximation for $g_\t{UL}$ can be obtained by assuming a median value $\hat{D}=0$, yielding the median constraint
\begin{equation}\label{eq:DCLmedian}
    D_\t{UL}^\text{med}=\sqrt{2} \text{Erf}^{-1} \left[ \text{CL}\right] \sigma_{\hat{D}} \equiv N_\t{CL}\sigma_{\hat{D}}
\end{equation}
and a $2\sigma$ ($\t{CL} = 0.95$, \textcolor{black}{$N_\t{CL}\approx 2$}) bound of \color{black}
\begin{equation}\label{eq:g_2sigma}
g^{(2\sigma)}\approx \frac{2\sqrt{\sigma_{\hat{D}}}}{a_0\Delta_{12}}
\end{equation}\color{black}

Assuming sufficient frequency resolution to resolve the DM lineshape ($\gamma_\t{DM}\tau_\t{p}>1$), and that detector noise (thermomechanical noise and shot noise) is constant over $\gamma_\t{DM}$, one can approximate \textcolor{black}{$\sigma_{\hat{S}^{\smash{(i)}}_a
}[\omega_k]\approx \sigma_{\hat{S}^{\smash{(i)}}_a
}[\omega_\t{DM}]\approx S_a^{(i)}[\omega_\t{DM}]/\sqrt{N_i}$} and
\begin{equation}
    \sigma^2_{\hat{D}}[\omega_{\t{DM}}]\approx\frac{\gamma_\t{DM}}{3\tau_\t{p}}\bigg( \sum_{i} \frac{N_i}{\big(S_a^{(i)}[\omega_\t{DM}]\big)^2}\bigg)^{-1}
\end{equation}
using Eq. \ref{eq:DMestimatorVariance} and the numerical approximation.
\begin{equation}
    \sum_{k}\left( G_\t{DM}^{\tau_\t{p}}[\omega_k]\right)^2\approx \tau_\t{p} \int_0^\infty G_\t{DM}^2[\omega]\frac{d\omega}{2\pi}\approx \frac{3\tau_\t{p}}{\gamma_\t{DM}}.
\end{equation}

Equation \ref{eq:g_2sigma} then yields
\begin{equation}
    g^{(2\sigma)}\approx \frac{2}{a_0\Delta_{12}}\left(\frac{\gamma_\t{DM}}{3\tau_\t{p}}\right)^{1/4}\bigg( \sum_{i} \frac{N_i}{\big(S_a^{(i)}[\omega_\t{DM}]\big)^2}\bigg)^{-1/4},
\end{equation}
which simplifies to 
\begin{equation}
    g^{(2\sigma)}\approx \frac{ \sqrt{S_a[\omega_\t{DM}]\gamma_\t{DM}}}{a_0\Delta_{12}}\left(\frac{8}{3}\right)^{1/4}\left(\frac{2}{\gamma_\t{DM}\tau}\right)^{1/4}
\end{equation}
for a single measurement $\hat{S}_a[\omega_k]$ with mean (fitted) value $S_a[\omega]$ and total measurement time $\tau = \tau_\t{p}N$.

Assuming the acceleration measurement is thermal noise limited $S_a [\omega_\t{DM}] = 4 k_B T \omega_\t{DM}/(m Q_\t{m})$ and defining $Q_\t{DM}\equiv \omega_\t{DM}/\gamma_\t{DM}$, it is straightforward to show that
\begin{equation}
    g^{(2\sigma)} = 2g^\t{(th)}\times (2/3)^{1/4}
\end{equation}
where $g^\t{(th)}$ is the thermal-noise-equivalent coupling strength given by Eq. 2 in the main text.  The extra factors of $2$ and $(2/3)^{1/4}\approx 0.9$ arise from the fact that $g^\t{(th)}$ is a $1\sigma$ constraint and the definition of $\gamma_\t{DM}$ for lineshape $G_\t{DM}[\omega]$,~respectively.

\color{black}
\subsubsection{Frequentist upper limit}

In a frequentist framework, $D_\t{UL}$ corresponds to the signal power that would produce a measurement outcome $\hat{D}$ (upon repeated experiments) with probability
\begin{equation}\label{eq:DUL_freq}
	\t{CL} = \int_{\hat{D}}^\infty \rho(\hat{D}'|D = D_\t{UL})d\hat{D}'
\end{equation}

Solving Eq. \ref{eq:DUL_freq} with the Gaussian sampling distribution given in Eq. \ref{eq:DDMgaussLikelihood} yields
\begin{equation}\label{eq:DUL_freq_2}
 D_\t{UL} = \hat{D}+\sqrt{2}\sigma_{\hat{D}}\t{Erf}^{-1}[2\t{CL}-1]
\end{equation}
and a median constraint
\begin{equation}
	D_\t{UL}^\t{med} = \sqrt{2}\sigma_{\hat{D}}\t{Erf}^{-1}[2\t{CL}-1]
\end{equation}
At $\t{CL} = 0.95$, this corresponds to $D_\t{UL}^\t{med} \approx 1.64\sigma_{\hat{D}}$ and an upper limit $g\propto \sqrt{D_\t{UL}}$ (Eq. \ref{eq:DCL}) which is $9.5\%$ smaller than obtained using the Bayesian approach ($D_\t{UL}^\t{med} \approx 2.0\sigma_{\hat{D}}$).
.
We note that \ref{eq:DUL_freq} does not enforce positivity $D>0$, such that $D_\t{UL}<0$ for sufficiently negative $\hat{D}$.  Positivity can be imposed by restricting solutions of Eq. S35 to $D_\t{UL}>0$ (e.g. via likelihood ratio ordering). This does not change $D_\t{UL}^\t{med}$.
\color{black}

\section{Experiment}

\subsection{Vibration isolation}
\begin{figure}[h!]
    \centering
\includegraphics[width=0.99\columnwidth,trim= 0in 0in 0in 0in]{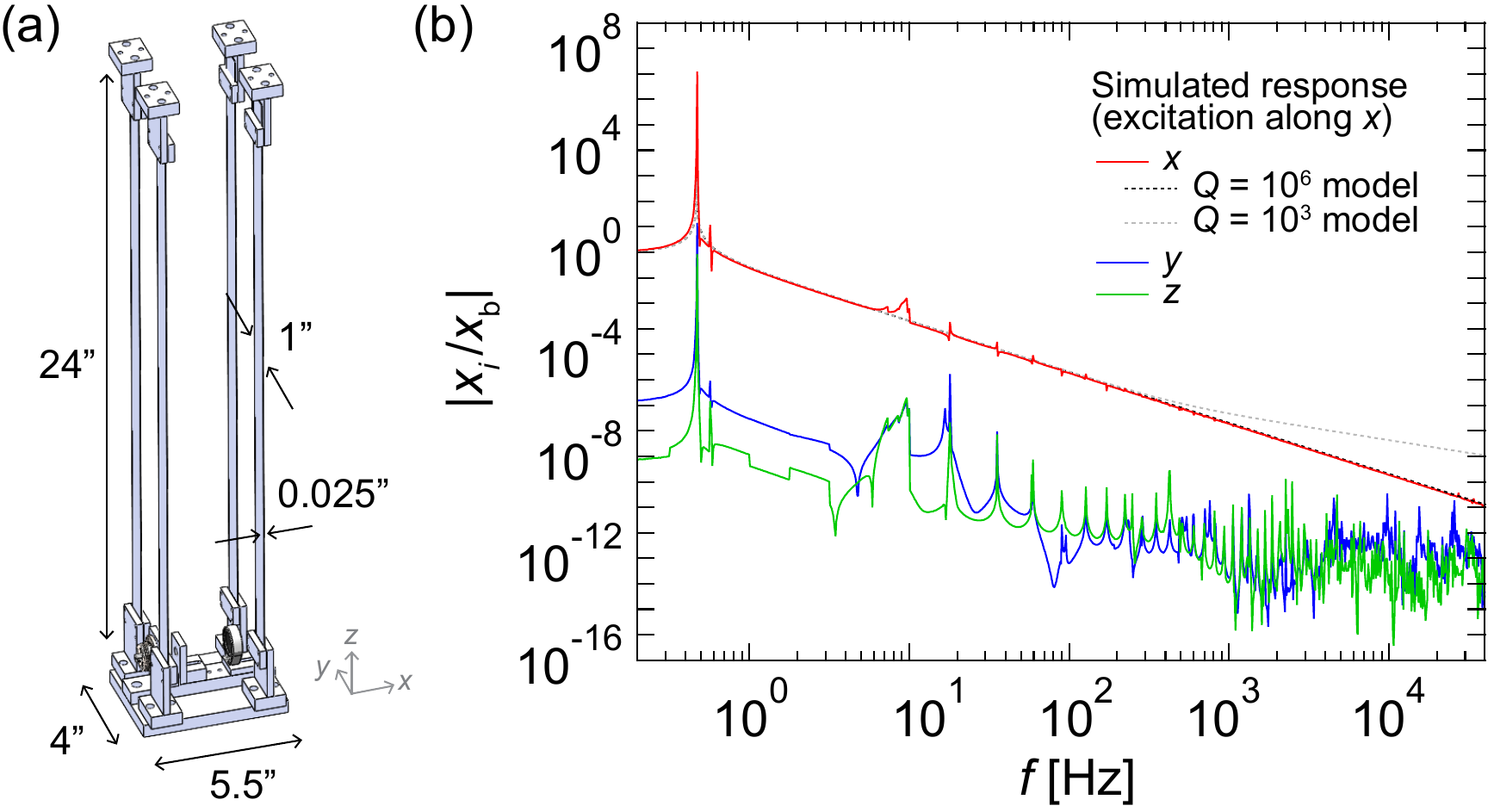}
    \caption{(a) Design of the vibration isolation system including the optical assembly. (b) Simulated transmissibility of the pendulum VIS along three directions ($x,y,z$).  The dual-membrane accelerometer is oriented along $x$ and gravity is oriented along $z$.}
    \label{fig:S2}
\end{figure}

Our vibration isolation system (VIS) is based on a similar system described in \cite{planz2023membrane}.  It consists of four 1 inch-wide (1 inch = 1" = 25.4 mm), 0.025 inch-thick, 24 inch-long Cu ribbons from which a miniature optical breadboard (also Cu) is suspended from the cryostat mixing stage, forming a $\sim1\;\t{Hz}$ pendulum.  Photographs of the VIS and breadboard are shown in Fig. 2(a) and 2(b) of the main text, respectively; a detailed design is shown in Fig. \ref{fig:S2}(a).  Finite element simulations of the fundamental longitudinal (along the detector axis) and lateral (transverse so the detector axis) pendulum modes are shown in main text Fig. 2(c), exhibiting resonance frequencies of $\omega_0 = 2\pi\times 0.5$ Hz and $2\pi\times 16$ Hz, respectively.  To generate these simulations, we use the Structural Mechanics module in COMSOL \cite{comsol2024,NISTdisclaimer} and assume a rectangular baseplate of mass of 1.6 kg and the dimensions shown in Fig. \ref{fig:S2}(a). A simulation of the VIS transmissibility $\mathcal{T}_{x_i x}[\omega]\equiv |x_i[\omega]/x_\t{b}[\omega]|$ along the longitudinal ($x_1 = x$), transverse ($x_2 = y$) and vertical ($x_3 = z$) directions is shown in Fig. \ref{fig:S2}(b), assuming base excitation $x_\t{b}$ (displacement of the mixing stage) along the longitudinal direction.  Characteristic $\mathcal{T}_{xx}[\omega\gg \omega_0]\propto (\omega_0/\omega)^2$ isolation is predicted in high-$Q$ limit, interspersed with  minor resonances corresponding to violin modes of the tensioned ribbon suspensions. Simulations of the fundamental longitudinal and transverse VIS modes are shown \mbox{in main text Fig. 4(e).}

\textcolor{black}{\subsection{Broadband displacement spectrum}}

\begin{figure}[b!]
    \centering
\includegraphics[width=0.9\columnwidth,trim= 0in 0in 0in 0in]{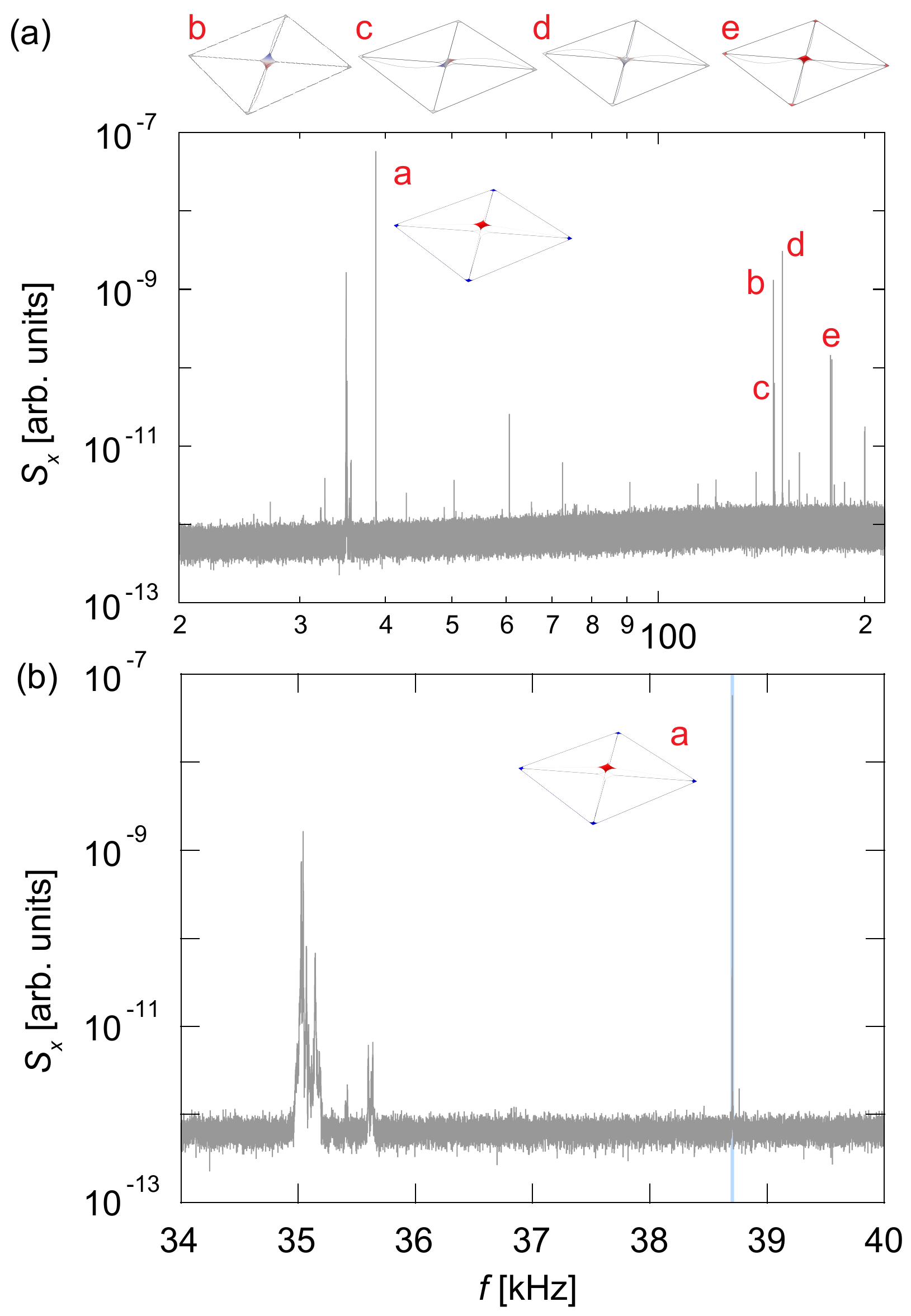}
    \caption{(a) A broadband displacement spectrum highlighting trampoline modes. (b) Zoomed in version of the same PSD showing spectral features between 35 to 36 kHz, in the vicinity of the (slightly damped) 39 kHz trampoline mode. The blue band corresponds to the extent over which the trampoline mode is scanned (see section III.D.)}\label{fig:SModes}
\end{figure}

\color{black}
We briefly address some of the spectral features observed at relatively high frequencies (inside the isolation band of the VIS), near or above the 39 kHz trampoline mode of interest.
Fig. \ref{fig:SModes}(a) shows a PSD estimate up to $\approx200$ kHz, taken using a high-bandwidth digitizer (note that this is different from the digitizer used to acquire the spectra used everywhere else in this work, which has a 100 kHz Nyquist frequency).
Most of the prominent peaks above the noise floor correspond to an assortment of trampoline modes. The 5 lowest-frequency modes are labeled in the spectrum; their frequencies agree well with predictions from COMSOL simulation. Displacement profiles of these modes are also shown.
A notable feature that is not explained by simulations of trampoline modes is the spectral structure between 35 to 36 kHz, as seen in the zoomed-in PSD in Fig. \ref{fig:SModes}(b). These peaks are low-$Q$ and unresponsive to a coherent radiation pressure drive acting locally on the trampoline pad, implying that they are likely frame modes of the chip (we observed features at similar frequencies in another experiment with a nearly identical trampoline \cite{pluchar2023thermal} at room temperature), or from the surrounding apparatus. The presence of these nearby features does not appear to affect the thermalization of the 39 kHz mode, as verified in the following section.
\color{black}

\subsection{Thermalization to cryogenic base temperature}\label{sec:Thermalization}

\begin{figure*}[t!]
		\vspace{-2mm}
		\centering  \includegraphics[width=1.65\columnwidth]{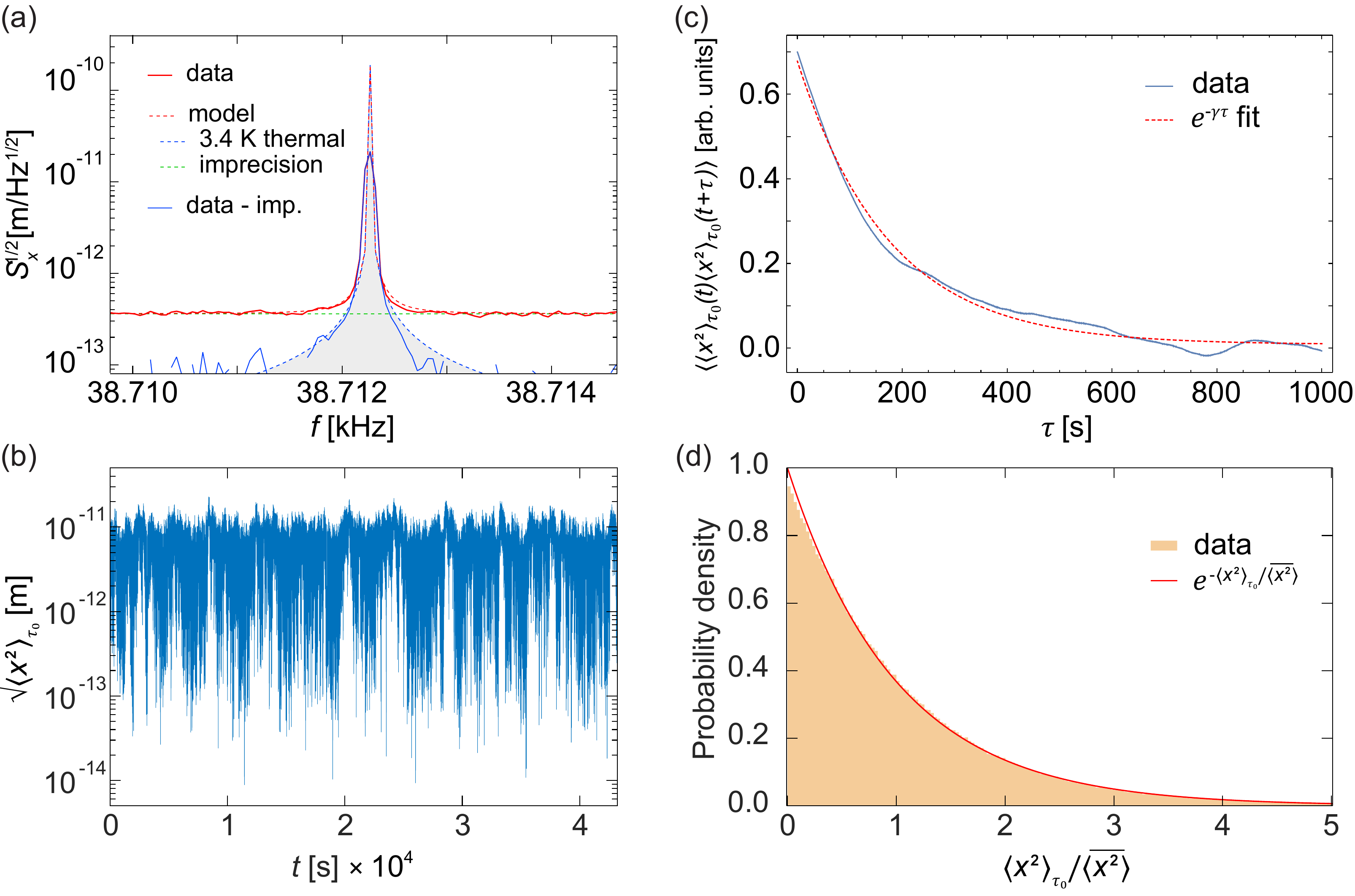}
\caption{\textbf{Thermalization to cryostat base temperature.} (a) Comparison of calibrated thermomechanical noise to cold-damped model, Eq.~\ref{eq:SI_Sthmodel}. (b) Simultaneously tracked instantaneous displacement power, plotted in root-mean-square units. (c) Autocorrelation of data in (b).  (d) Histogram of instantaneous displacement power compared to a Boltzmann distribution.  }
		\label{fig:S3}
	\end{figure*}

In Fig. S3, we present three measurements evidencing thermalization of the $\omega_\text{m}=2\pi\times 39$ kHz fundamental trampoline mode to the $3.4$ K base temperature of the LD-4K cryostat (as reported by the thermistor on the mixing stage). These include: (1) [Fig. \ref{fig:S3}(a)] a displacement PSD estimate compared to a thermal noise model; (2) [Fig. \ref{fig:S3}(b,d)] histograms of the real-time displacement power compared to a Boltzmann distribution, and (3) [Fig. \ref{fig:S3}(b,c)] energy-autocorrelation measurements used to infer the loaded mechanical damping rate, and thereby the bath temperature, from a cold-damping model.  All measurements were carried out with a weak 850 nm probe beam ($P_\text{out}=20 \;\upmu$W) and no feedback beam, to minimize photothermal heating (see Sec. III.C).  Photosignals were calibrated in displacement units using the side of the fringe method \cite{chowdhury2023membrane}.

To compare the apparent bath temperature $T_0$ of the trampoline to the cryostat base temperature, the displacement PSD estimate in Fig. \ref{fig:S3}(a) is overlaid with a thermal noise model 
\begin{subequations}\label{eq:SI_Sthmodel}\begin{align}
    S_x[\omega] &=|\chi[\omega,\gamma]|^2  4 k_B T_0 m \gamma_0 + \frac{|\chi[\omega,\gamma]|^2}{|\chi[\omega,\gamma_0]|^2}S_x^\t{imp}\\&\approx \frac{ 4 k_B T_0 \gamma_0/(\gamma^2 m \omega_\t{m}^2)}{1+4(\omega-\omega_\t{m})^2/\gamma^2}  + S_x^\t{imp}
\end{align}\end{subequations} where $m$ is the effective mass, $\gamma_0$ is the intrinsic mechanical damping rate and $\gamma$ is the loaded damping rate (accounting for photothermal backaction). Using $T_0 = 3.4\;\t{K}$, we find that data and model are good agreement in the wings of the thermal noise peak, $\omega-\omega_\t{m}\gg\gamma$, where $S_x[\omega]\approx  k_B T_0 \gamma_0 /(m\omega_\t{m}^2(\omega-\omega_\t{m})^2) + S_x^\t{imp}$ depends only on $T_0$, $\gamma_0$ and $m$.  We take this as indirect evidence that $T_0\approx 3.4$ K, subject to uncertainty in $m\approx 12$ ng, which is obtained using a finite element model for the photolithographically defined trampoline dimensions \cite{chowdhury2023membrane}, and $\gamma_0 = 2\pi \times 0.7\;\t{mHz}$, which we infer from the ringdown measurement shown in Fig. 2(e) of the main text.

As shown in Fig. \ref{fig:S3}(b), simultaneous with the PSD estimate in Fig. S3(a), we tracked the instantaneous displacement  power $\langle x^2 \rangle_{\tau_0}$ \cite{gieseler2014dynamic} (corresponding to the area beneath the displacement noise peak integrated over a bandwidth $\tau_0^{-1}$ such that $\omega_\t{m}\gg\tau_0^{-1} \gg \gamma$) for 12 hours using a lock-in amplifier (Zurich Instruments MFLI). A histogram of $\{\langle x^2\rangle_{\tau_0}\}$ is compared to the Boltzmann probability distributions $p(\langle x^2\rangle) \propto e^{-\langle x^2\rangle /\overline{\langle x^2\rangle}}$ in Fig. \ref{fig:S3}(d), where $\overline{\langle x^2\rangle}$ is the ensemble average power.  Quantitative agreement provides further evidence that the oscillator is in a thermal state. From the ensemble average power, the effective temperature of the mode is inferred to be $T_\t{mode} = \overline{\langle x^2\rangle}/(k_\t{B} m \omega_\t{m}^2) = 2.4$ K.

To infer the bath temperature from the modal temperature, we use the cold damping model \cite{aspelmeyer2014cavity}
\begin{equation}
T_\t{mode} = T_0\gamma_0/\gamma
\end{equation}
and extract the loaded damping rate from an exponential fit to the instantaneous power autocorrelation $\langle \langle x ^2\rangle_{\tau_0}(t) \langle x ^2\rangle_{\tau_0}(t+\tau) \rangle \propto e^{-\gamma \tau}$ \cite{davenport1958introduction,zheng2020room}, as shown in Fig. \ref{fig:S3}(c).  From the fitted value $\gamma = 2\pi\times 1\;\t{mHz}$, we infer  $T_0 = 3.5$ K, in good agreement with the temperature reported by the mixing stage thermistor. 




\subsection{Photothermal heating and frequency tuning}

\begin{figure}[h!]
    \centering
\includegraphics[width=0.9\columnwidth,trim= 0in 0in 0in 0in]{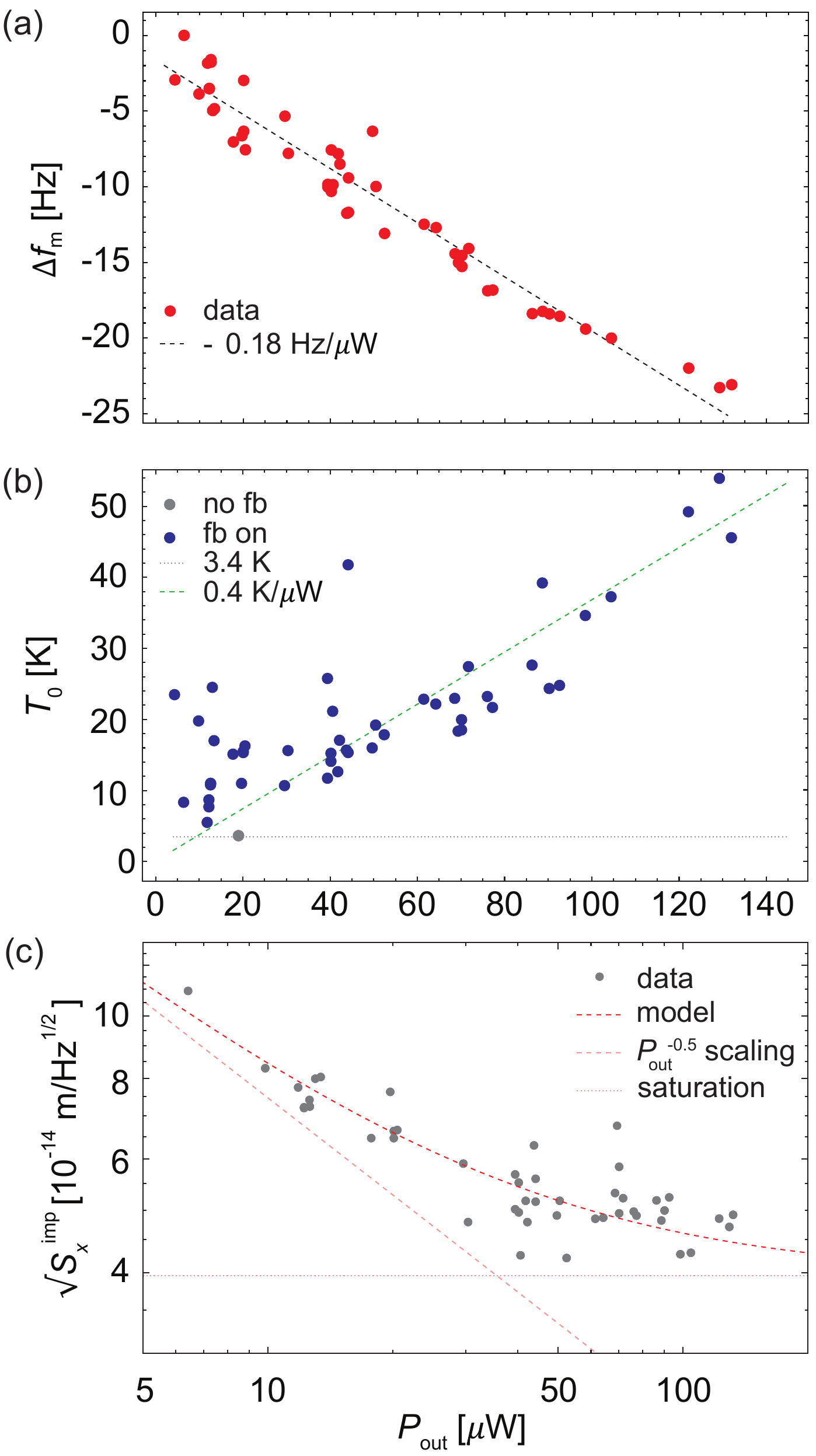}
    \caption{Photothermal tuning by varying probe power. (a) Resonance frequency shift (red points) of the 39 kHz mode vs. power. We infer 0.18 Hz/$\upmu$W (fit: dashed line) tuning rate at side-of-fringe detuning. (b) Concomitant photothermal heating (blue points) at 0.37 K/$\upmu$W (green dashed line). Minimum temperature (feedback laser off, corresponding to Fig. \ref{fig:S3}) is plotted for reference (gray point). (c) Reduction in local imprecision noise as probe power is increased until the classical noise floor of the probe is saturated.}\label{fig:S4}
\end{figure}

For measurements shown in Fig. 3 of the main text, the trampoline is subject to photothermal heating (elevation of the bath temperature $T_0$) due to absorption of light from the 850 nm probe beam and 650 nm feedback beam \textcolor{black}{(see footnote [31] in the main text).}
We infer this heating by fitting the feedback-damped, spectrally resolved PSD to the thermal noise model in Eq. \ref{eq:SI_Sthmodel}, using $T_0$ as an independent variable.

Photothermal heating is used to effect a frequency scan, as shown in Fig. \ref{fig:S4}(a) (same as inset in Fig. 3(d) of the main text) by varying the probe power, $P_\t{out}$.  For this scan, we fixed the mean power in the feedback laser to $P_\t{out,fb}\approx 10 \;\upmu$W and varied $P_\t{out}$ from $5\;\upmu\t{W}$ to $132\;\upmu\t{W}$.  
A plot of bath temperature (using spectral fitting) versus probe power is shown in Fig. \ref{fig:S4}(b), exhibiting linear heating with a rate of $d T_0/d P_\t{out}\approx 0.4\;\t{K}/\upmu\t{W}$. For the smallest probe powers, heating is dominated by the feedback beam and the bath temperature was determined to be $T_0 = 8.3$ K (compared to 3.5 K with $P_\t{out,fb}=0$, gray point in Fig. \ref{fig:S4}(b))---this sets the minimum thermal acceleration noise for the data in main text Fig. 3.

Finally, we note that the photothermal frequency shift was observed to have a non-trivial dependence on probe-cavity detuning, suggesting a form of cavity-assisted photothermal backaction \cite{metzger2004cavity}. 
For the UDM search, we chose to fix the detuning to the point of maximum displacement sensitivity (i.e. maximum slope) on the cavity fringe, which coincidentally was found to maximize the photothermal damping.

\subsection{Frequency tuning via radiation pressure: Outlook}
\color{black}

While we leveraged photothermal effects for resonance tuning, future experiments---especially at dilution refrigerator temperatures ($\lesssim 100\;\t{mK}$)---will require alternative tuning mechanisms to prevent sensitivity loss due to modal heating. 

The approach we envision involves integrating the membrane test mass into a high finesse optical cavity and exploiting the  optical spring shift due to radiation pressure dynamical backaction \cite{aspelmeyer2014cavity,rohse2020cavity,planz2023membrane,pluchar2023thermal}.  For example, a 40 kHz trampoline with identical dimensions as ours, integrated into a $500\;\mu\t
m$-long, $\mathcal{F}=10^3$ cavity and probed with $100\;\mu\t{W}$ of $\lambda\approx850\;\t{nm}$ light at an optimal detuning, would experience a frequency shift of $\Delta f_\t{m} = -16$ kHz (here we use parameters from the ``trampoline-in-the-middle'' experiment described in  \cite{pluchar2020towards}). 

As discussed in Sec. \ref{Sec:FutureMembranes}, we envision using larger membranes with higher mass and thermal conductivity (mitigating photothermal heating) in future experiments. 
For the $1.5\;\mu\t{g}$, $7.5\;\t{kHz}$ ``sail'' membrane described in Table \ref{table:sail} and shown in Fig. \ref{fig:S7}(b), the optical spring produced by the same cavity parameters as above would be $\Delta f_\t{m}\approx-0.7\;\t{kHz}$. 

As discussed in \cite{manley2021searching,brady2023entanglement}, our long-term goal is to combine frequency scanning with multi-mode readout and an array of devices. This allows for increased bandwidth in addition to sensitivity enhancement from integrating the signal in a quantum-coherent way across multiple sensors. 

\color{black}

\subsection{Displacement readout shot noise}

\textcolor{black}{Increasing the probe power to effect frequency tuning concomitantly leads to increased displacement sensitivity (hence, bandwidth---see main text Fig. 3(c), and \cite{chowdhury2023membrane}) by shot noise reduction. In Fig. \ref{fig:S4}(c), we plot the mean displacement noise floor as a function of transmitted power $P_\t{out}$ over a 1 kHz window from 38.2 to 39.2 kHz (obtained from a fit to the data using $S_x^\t{imp}$ as an independent variable). The $P_\t{out}^{-0.5}$ noise scaling shows that the readout was shot-noise limited up to $P_\t{out}\approx100\;\upmu W$. At higher powers, we observe a technical noise floor of $4\times10^{-14}\;\t{m/\sqrt{\t{Hz}}}$. These measurements were carried out with active intensity noise stabilization \cite{pluchar2025quantum}. In future experiments, we anticipate that increased intensity feedback gain (our current limitation) and an active frequency-stabilization will enable lower technical noise.} 


\begin{figure}[b!]
    \centering
\includegraphics[width=0.92\columnwidth,trim= 0in 0in 0in 0in]{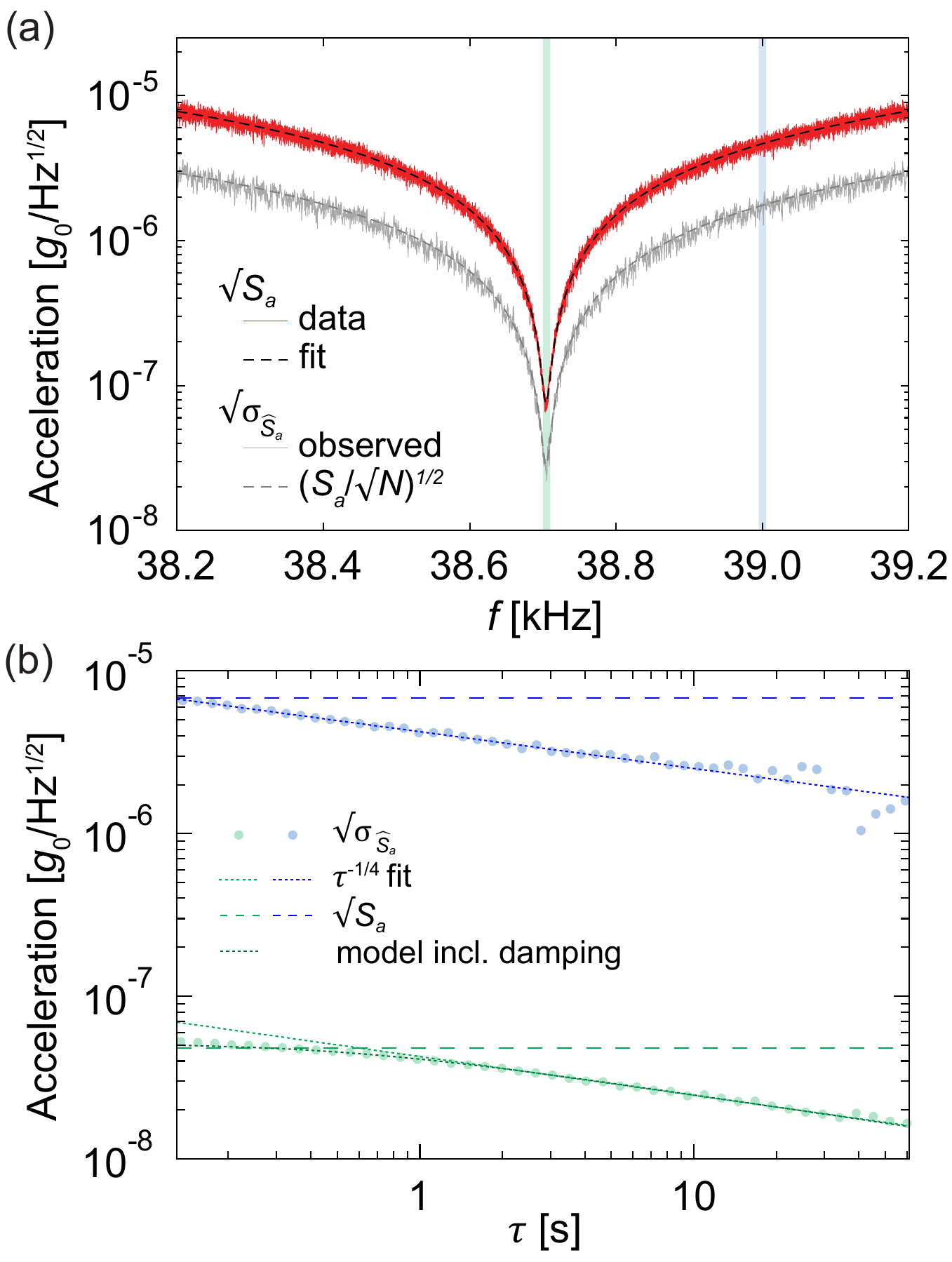}
    \caption{Noise averaging example. (a) PSD estimate (red) after $N = 50$ averages compared to the standard deviation over a moving 1 Hz window (light gray) in the estimate. The dashed gray line represents the expected noise level calculated from the fit (dashed black) after 50 averages. (b) Noise reduction in the PSD estimate as a function of averaging time $\tau$. Green (blue) data corresponds to thermal (imprecision) noise resolution on (off) resonance, highlighting $\tau^{-1/4}\propto N^{-1/4}$ scaling. On resonance, this scaling is attained after $\tau\approx 1/\gamma$ as shown in the model overlay including damping.}
    \label{fig:S5}
\end{figure}

\subsection{Noise averaging: An example}
As discussed in the main text (Eq. 3) and Sec. II, the standard deviation of the acceleration PSD estimate $\sigma_{\hat{S}_a}[\omega] = S_a[\omega]/\sqrt{N}$ reduces as a function of total averaging time $\tau = N\tau_\t{p}$, where $\tau_\t{p}$ is the duration of a single time trace (used to produce a periodogram $\hat{S}_a^{\tau_\t{p}}$), $N$ is the number of time traces (periodograms) averaged to produce $\hat{S}_a$, and $S_a$ is the true PSD.
Fig. \ref{fig:S5}(a) illustrates this scaling for a single representative measurement $\hat{S}_a[\omega]$ with $N = 50$. The gray trace is the moving standard deviation $\smash{\sqrt{\sigma_{\hat{S}_a}[\omega]}}$ over a 1 Hz window (chosen arbitrarily). The black line is the expected value, with $S_a$ inferred from a least-squares fit. Good agreement is observed.
Fig. \ref{fig:S5}(b) further illustrates the scaling $\smash{\sqrt{\sigma_{\hat{S}_a}}}\propto \tau^{-1/4}$ by recording a sequence of instantaneous power measurements centered at frequency $\omega_0$, $\langle x^2 \rangle_{\tau_0}(\omega_0)$ [cf. Fig. S3(b), with $\omega_0 = \omega_\t{m}$], using the approximation
\begin{equation}
\langle x^2 \rangle_{\tau_0}(\omega_0)= \int_{\omega_0-\tfrac{\pi}{\tau_0}}^{\omega_0+\tfrac{\pi}{\tau_0}}\hat{S}_x^{\tau_0}[\omega]\tfrac{d\omega}{2\pi}\approx \hat{S}_a^{\tau_0}[\omega_0]\int_{\omega_0-\tfrac{\pi}{\tau_0}}^{\omega_0+\tfrac{\pi}{\tau_0}}|\chi[\omega]|^2\tfrac{d\omega}{2\pi}
\end{equation}
and computing the standard deviation of the sequence after binning into non-overlapping intervals of length $\tau=N\tau_0$.
The data in Fig. \ref{fig:S5}(b) confirms $\tau^{-1/4}$ scaling for both on-resonance ($\omega_0 = \omega_\t{m}$) and off-resonance ($\omega_0 = \omega_\t{m}+2\pi\cdot300\;\t{Hz}$) PSD estimates with $\tau_0 = 0.1\;\t{s}\ll\gamma^{-1}$, yielding the expected 2.5-fold increase in resolution after $\tau = 50$ seconds of averaging. In the former case, $\tau^{-1/4}$ scaling is reached for $\tau>1/\gamma$ (dark green line). The observed behavior is consistent with demonstrations in \cite{gavartin2012hybrid,chowdhury2023membrane} that, for a thermal-noise dominated acceleration background, cold-damping can reduce the time necessary to achieve a target noise resolution by a factor of $\gamma/\gamma_0$ ($\approx 900$ in this case), where $\gamma$ is cold damping rate.

\begin{figure}[h!]
    \centering
\includegraphics[width=0.85\columnwidth,trim= 0in 0in 0in 0in]{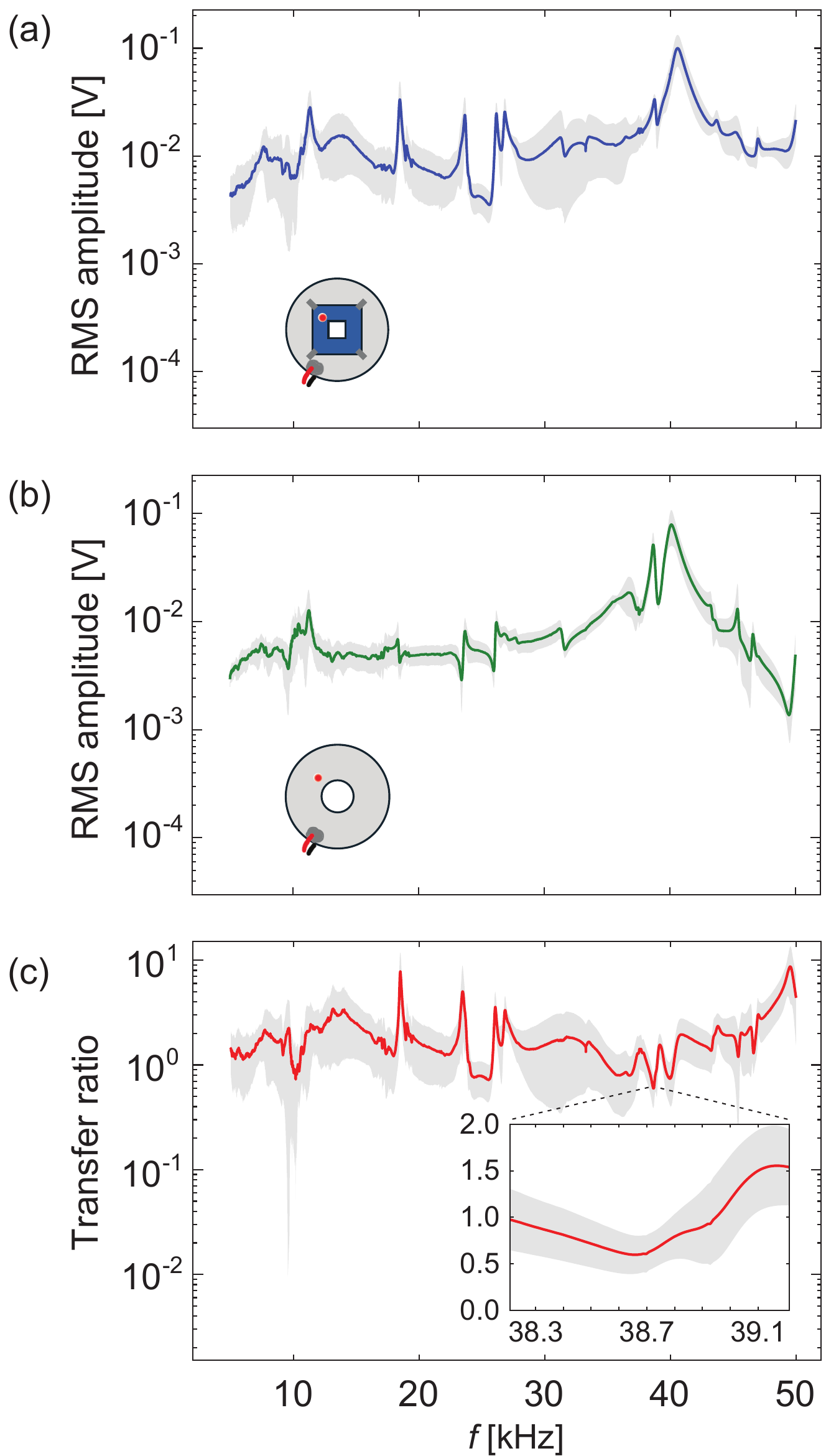}
    \caption{Combined measurements of displacement when measured on (a) chip on piezo, and (b) piezo directly. (c) Inferred wide-band and near-resonant (inset) transfer function.}
    \label{fig:S6}
\end{figure}

\color{black}\subsection{Fractional charge ratio $\Delta_{12}$} 

The fractional charge ratio used in our detector model is $\Delta_{12} = (Z/A)_\t{Cu} - (Z/A)_\t{Si_3 N_4}= 0.04$, where $(Z/A)_\t{Cu} = 0.46$ and $(Z/A)_\t{Si_3 N_4} = 0.50$ are the charge ratio of Cu and Si$_3$N$_4$, respectively.  This model treats the Cu base plate and the trampoline as a heterogeneous mechanical dimer \cite{manley2021searching}, and is based on several assumptions: (1) we neglect contributions to the base plate charge ratio from the lens (Bk7 glass), lens mount (Al) and FiberPort (steel), since they constitute $<2\%$ of the base plate mass; (2) we neglect the motion of the cryostat, since it is decoupled from the base plate at frequencies much larger than the $f_\t{VIS}\sim 1\;\t{Hz}$ vibration isolation corner frequency (at the trampoline frequency $f_\t{m}\approx 40$ kHz, the base plate approximates free-fall), (3) we neglect the material inhomogeneity (Si and Si$_3$N$_4$) of the double-membrane accelerometer, since $Z/A_\t{Si}-Z/A_\t{Si_3 N_4} < 0.001$; (4) we assume that the double-membrane accelerometer is rigidly attached to the Cu plate, such that trampoline and square membrane experience an effective base acceleration $a_\t{Cu}-a_\t{Si_3 N_4}$, and (5) we assume that the square membrane is much stiffer than the trampoline, such that it co-accelerates with the Cu base at frequencies $f\approx f_\t{m}$. Assumption 5 was addressed in \cite{chowdhury2023membrane}, and 4 is explored in the next section.
\color{black}

\subsection{Chip rigidity and transfer function}

The differential acceleration between the two membranes may differ from the lumped-mass model (main text Fig. 1) if the chip is not perfectly rigid. To investigate this discrepancy, we measured the transfer function between the base to which the chip is attached (glued on its four corners) to the inside corner of the chip where the membrane is suspended, emulating the mounting geometry of the cryogenic device chip.
The setup consists of an auxiliary device chip--which has the same dimensions as the chip used for the UDM search, but no suspended membranes--glued on four corners to a ring-shaped piezoelectric plate transducer (PZT).  
A Michelson interferometer was used to probe the displacement at two positions: (a) near an inside corner of the chip, and (b) directly on the PZT close to the outer corner of the chip where it is glued.
The experiment was performed in atmospheric pressure at 300 K, under the assumption that the chip response is similar when operated in a cryogenic, high vacuum environment. (As a future upgrade, we plan to install a PZT directly onto the sample stage to enable cryogenic chip response measurements.)

We define the chip's transfer function as 
\begin{equation}\label{eq:transferratio}
    r_\text{chip}(\omega) \equiv \frac{|\tilde{a}_\text{eff}(\omega)|}{|\tilde{a}_\text{base}(\omega)|} = \frac{|\tilde{x}_\text{eff}(\omega)|}{|\tilde{x}_\text{base}(\omega)|}
\end{equation}
where $\tilde{a}_\text{eff}(\omega)$ [$\tilde{x}_\text{eff}(\omega)$] and $\tilde{a}_\text{base}(\omega)$ [$\tilde{x}_\text{base}(\omega)$] are the effective acceleration [displacement] $\tilde{a}_\text{eff}(\omega)$ [$\tilde{x}_\text{eff}(\omega)$] of the trampoline test mass and base, respectively, and $\omega$ is the excitation frequency. As shown in the inset to Fig. \ref{fig:S6}(a), $\tilde{x}_\text{eff}(\omega)$ is measured at the inside corner of the Si chip where the trampoline would be suspended (red point in panel a).  Separately, $\tilde{x}_\text{base}(\omega)$ is measured directly on the PZT base (with chip absent) near the inside boundary of the PZT where the chip would be in contact with the base (red point in panel b).

Response measurements are summarized in Figs. \ref{fig:S6}. Panels (a,b) shows root mean square (RMS) displacement amplitudes $|\tilde{x}_\text{eff}(\omega)|$ and $|\tilde{x}_\text{base}(\omega)|$ at PZT excitation frequency $f = \omega/(2\pi)$, in units of voltage produced by the interferometer photodetector. (The interferometer power balance (including absolute signal and local oscillator powers) and phase, photodetector gain, and PZT excitation power were kept constant across all measurements.) The shaded region in each plot shows the standard deviation due to multiple experimental runs. The corresponding estimate of $r_\text{chip} (\omega)$ obtained via Eq. \ref{eq:transferratio}, is plotted in panel (c). We find that in the 1 kHz window of our UDM search analysis (main text Fig. 4), $r_\text{chip}(\omega)\in[0.5,1]$, indicating the chip is approximately rigid at these frequencies.

In our UDM search analysis (Fig. 4 of the main text), we have limited our search bandwidth to a 1 kHz window around the trampoline resonance where the chip transfer function is near unity. To compensate for chip compliance, we make the substitution $S_a(\omega) \rightarrow S_a(\omega)/r_\text{chip}^2(\omega)$, yielding $\sigma_{\hat{D}}\rightarrow \sigma_{\hat{D}}/r_\text{chip}^2$. The red trace in Fig. 4(c) of the main text highlights this correction using the emperical $r_\text{chip}$ in Fig. \ref{fig:S6}(c), while the gray trace assumes $r_\text{chip} = 1$.

\vspace{-1mm}
\section{Outlook}\label{Sec:FutureMembranes}

\subsection{Optimized ``Gen 2" membrane accelerometers}

\begin{figure}[ht!]
    \centering
\includegraphics[width=0.76\columnwidth,trim= 0in 0in 0in 0in]{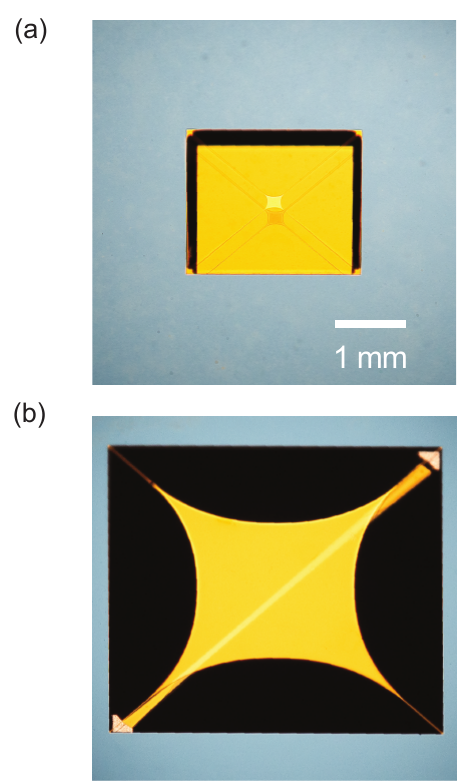}

    \caption{A comparison of (a) the trampoline test mass used in this paper and (b) a larger trampoline (``sail'') with Bayesian-optimized dimensions to maximize $m\times Q$ product \cite{hyatt2025ultrahigh}. Both devices are dual-membrane cavities: a square membrane [diagonal beam] constitutes the rigid back mirror in (a) [(b)]. Images are to scale.}
    \label{fig:S7}
\end{figure}

In future experiments, we plan to use more massive test masses designed to suppress thermal acceleration noise by maximizing the quantity $m\times Q_\t{m}/f_\t{m}$.
To increase $m$ without sacrificing $Q_\t{m}$ or increasing $f_\t{m}$, we have performed Bayesian optimization to enlarge the side length of the trampoline pad [Fig. \ref{fig:S7}(a)] from $200\;\upmu\t{m}$ to $2.5\;\t{mm}$, thereby realizing a ``sail''-like test mass [\ref{fig:S7}(b)]. The sail---suspended from a 5 mm window as opposed to 2.5 mm for the trampoline---has $m\approx 1.5\;\upmu\t{g}$, $\sim\!100$ times more massive the trampoline.
The sail has approximately the same thickness ($\lesssim 100\;\t{nm}$) as the trampoline and slightly wider tethers ($10\;\upmu\t{m}$ vs. $4\;\upmu\t{m}$); finite element simulations predict $Q_\t{m}\approx 7\times10^7$ using a dissipation dilution model \cite{sadeghi2019influence,fedorov2019generalized}. To date, however, we have observed $Q_\t{m}=2\times10^6$ experimentally at 300 K and 4 K.

\begin{table}[b!]\label{table:sail}
    \begin{ruledtabular}
        \begin{tabular}{ l c c c c c }
         Test mass & $m\;\t{[\upmu g]}$ & $f_\t{m}\;\t{[kHz]}$ & $Q_\t{m}\;[10^7]$ & \multicolumn{2}{c}{$\sqrt{S_a^\t{th}}\;[\t{n}g_0/\t{Hz}^{1/2}]$}\\
           & & & &\ 4 K &\ 10 mK \\
         \colrule
          Trampoline & $1.2\times10^{-2}$ & 38.7 & 6 & 28 & 1\\
          Sail & 1.5 & 7.3 & & &\\
          $\ \ $Empirical & & & 0.2 & 6 & 0.3\\
          $\ \ $Predicted & & & 7 & 1 & 0.05\\
          \multirow{2}{7 em}{10 cm square proposed in \cite{manley2021searching}} & $5.4\times10^3$& 3.8& $10^2$ & $3\times10^{-3}$ & $2\times10^{-4}$\\
          & & & &\\
        \end{tabular}        
    \end{ruledtabular}
    \caption{Table S1. Mass, resonance frequency, quality factor, and predicted thermal acceleration noise on resonance for trampoline (used in this work), sail (ongoing work), and ideal \cite{manley2021searching} test masses at 4 K and 100 mK (assuming thermalization).}
    \label{tableS1}
\end{table}

Predictions for the sail's 
acceleration sensitivity $\sqrt{S_a^\t{th} [\omega_0]}$ at 4 K and 10 mK (dilution refrigerator) temperatures are summarized in Table \ref{tableS1}---for both predicted as well as empirically observed $Q_m$---and compared against the trampoline. Best estimates predict a 30-times sensitivity-enhancement vis-\`a-vis the trampoline, potentially reaching as low as $0.05\;\t{n}g_0/\t{Hz}^{1/2}$ at 10 mK. Realizing a novel sensitivity---proposed in Ref. \cite{manley2021searching} using an idealized $\sim\!10$ cm square membrane---would require a further 1000-fold improvement in acceleration sensitivity at $\sim 1\text{-}10\;\t{kHz}$; our current efforts focus on releasing larger sails and exploring ways to mass-load the devices.

\vspace{-3mm}\subsection{Comparison of Optomechanical Accelerometers as Ultralight Dark Matter Detectors}

In Table 1 and Fig. 5, we compare various optomechanical accelerometers and their projected performance as B-L UDM detectors, assuming thermal-noise-limited operation at temperatures as low as 10 mK. Also included are magnetically levitated systems employing electromechanical readout, including the POLONAISE \cite{amaral2025first} UDM detector.  These include:

\begin{enumerate}
    \item A 22 mg, 20 cm square Si$_3$N$_4$ membrane as originally proposed in~\cite{manley2021searching}.

    \item A 0.5 mg, 10 cm version of the membrane in~\cite{manley2021searching}.
    
    \item The 12 ng Si$_3$N$_4$ trampoline reported in this work.
    
    \item A  $1\;\upmu$g, centimeter-scale ``sail'' trampoline under exploration as the second generation device for our experiment. See Sec. IV.A for preliminary results.

    \item The $0.4$ mg permanent magnet levitated in a superconducting trap, used in the first generation POLONAISE UDM experiment \cite{amaral2025first}. The acceleration noise reported in \cite{amaral2025first} is $100$ times higher than projected, highlighting challenging vibration isolation requirements.

    \item A $1.8$ mg diamagnet levitated in a permanent magnet trap, 
    reported in \cite{xiong2025achievement}. An alternative to superconducting levitated setups, this system enables low eddy current damping in a levitated diamagnetic platform. 

    \item A $5.6\;\upmu\t{g}$ superconducting test mass levitated in a static magnetic trap (superconducting coil) \cite{hofer2023high}.
     An advantage of this magnetic levitation platform is suppression of dissipation from hysteresis and eddy currents.
    
    \item A 0.3 mg, lithographically defined torsion pendulum reported in \cite{condos2025ultralow}, formed by suspending a Si microchip from a Si$_3$N$_4$ nanoribbon.  An advantage of this platform is its arrayability and potential enhanced vibration isolation if heterogenously mass-loaded \cite{sun2025differential,manley2024microscale}.

    \item A 95 mg, flipchip mass-loaded Si$_3$N$_4$ trampoline recently reported in Ref. \cite{bawden2025precision}.  An advantage of this platform is the potential for heterogenous integration of test masses with high neutron density (e.g., Pt in \cite{bawden2025precision}).

    \item A 10 mg, dielectric-mirror-loaded Si$_3$N$_4$ trampoline reported in Ref. \cite{zhou2021broadband}.  An advantage of this platform is compatibility with integration in to a high finesse cavity, extending thermal-noise-limited bandwidth.  

    \item A 17 mg, photonic-integrated,  bulk Si accelerometer recently reported in \cite{ge2025towards}. An advantage of this platform is its compatibility with proposals for cryogenic  sensor arrays probed with distributed squeezed light \cite{brady2023entanglement}.

    \item A 2.6 g bulk fused silica accelerometer developed for advanced seismic monitoring and space-based geodesy \cite{hines2022optomechanical}. An advantage of this platform is its technical readiness for satellite deployment \cite{nelson2024six}, yielding the possibility of long-baseline UDM detector arrays \cite{derevianko2018detecting}.\\

   \item A 1.1 mg bulk fused silica accelerometer, with a test mass similar in design to device 12 above but with a  higher (0.7 kHz) resonance frequency. An advantage of this platform is that it is fully monolithic and enables a high $mQ_\t{m}$ product at relatively high frequencies \cite{guzman2018compact}. 
    
\onecolumngrid
\vfill
\vspace{0mm}
\begin{center}
  \begin{threeparttable}
    \setlength{\tabcolsep}{2pt}       
    \renewcommand{\arraystretch}{1.2} 
   \begin{tabular*}{\linewidth}{@{\extracolsep{\fill}} lcccccccc}
      \toprule
      \toprule
      Test mass & $m\;\t{[\upmu g]}$ & $\omega_\t{m}/2\pi\;\t{[Hz]}$ & \multicolumn{3}{c}{$Q_\t{m}\;[10^6]$} &
      \multicolumn{3}{c}{$\sqrt{S_a^\t{th}}\;[\t{n}g_0/\t{Hz}^{1/2}]$}\\
      & & & 300 K & 4 K & 10 mK\tnote{(a)} & 300 K & 4 K & 10 mK\tnote{(a)} \\
      \midrule
      1. 20 cm membrane \cite{manley2021searching} & $2.2\times10^4$ & $2\times10^{3}$ &  & & $10^3$ & & & $6\times10^{-5}$\\
      2. 10 cm membrane \cite{manley2021searching} & $5.4\times10^3$ & $3.8\times10^{3}$ &  & & $10^3$ & & & 
      $2\times10^{-4}$\\
      3. Trampoline [this work] & $1.2\times10^{-2}$ & $39.7\times10^3$ & $11$ & $60$ & $60$\tnote{(b)} & $6\times10^2$ & $28$ & 
      $1$\tnote{(b)}\\
      4. ``Sail'' resonator (Sec. IV.A) & $1.2$ & $7.3\times10^3$ & $2$ & $2$ & $70$\tnote{(c)} & $50$ & $6$\tnote{(b)} & 
      $0.3$\tnote{(c)}\\
      5. Levitated permanent magnet \cite{amaral2025first}& $4.3\times10^2$ & $27$ &  &  & $9.3$\tnote{(b)} &  &  & $5\times10^{-4}$\tnote{(b)}\\
      6. Levitated diamagnet \cite{xiong2025achievement} & $1.8\times10^3$ & $18$ & $0.1$ &  & $1$ & $0.3$ &  & $6\times10^{-4}$\tnote{(b)}\\
      7. Levitated superconductor \cite{hofer2023high} & $5.6$ & $240$ &  &  & $26$\tnote{(b)} &  &  & $8\times10^{-3}$\tnote{(b)}\\
      8. Mass-loaded micro-pendulum \cite{condos2025ultralow}& $3.3\times10^2$ & $98$ & 5 & & $25$ & $0.4$ &  & $9\times10^{-4}$\\
      9. Mass-loaded trampoline (flipchip) \cite{bawden2025precision}& $9.5\times10^4$ & $117$ & $2\times10^{-3}$\tnote{(d)} &  & $10^{-2}$ & 0.8 &  & $2\times10^{-3}$\\
      10. Mass-loaded trampoline (integrated) \cite{zhou2021broadband}& $1.0\times10^4$ & $9.9\times10^3$ & $1\times10^{-4}$\tnote{(d)} &  & $1$ & 100 &  &$6\times10^{-3}$\\
      11. Photonic-integrated bulk silicon \cite{ge2025towards}& $1.7\times10^4$ & $93\times10^3$ & $5\times10^{-5}$\tnote{(d)} &  & $10^{-2}$ & $3\times10^{2}$ &  & $0.1$\\
      12. Bulk fused silica \cite{hines2022optomechanical} & $2.6\times10^6$ & $4.7$ & $0.5$ &  & $0.5$\tnote{(b)} & $2\times10^{-3}$ &  & $1\times10^{-5}$\tnote{(b)}\\
      13. Bulk fused silica \cite{guzman2018compact} & $1.1\times10^3$ & $690$ & $0.5$ &  & $1$ & $0.1$ &  & $5\times10^{-4}$\\
      \bottomrule
      \bottomrule
    \end{tabular*}
    \begin{tablenotes}[flushleft]
\footnotesize
\item (a) projection \quad
(b) using best measured $Q_m$ \quad
(c) using $Q_m$ from simulation \quad
(d) measured in air
\end{tablenotes}
  \end{threeparttable}
  \refstepcounter{table} 

  \parbox{1\textwidth}{\small \text{TABLE~\thetable.} Comparison of optomechanical accelerometer test masses and projected thermal noise at cryogenic temperatures.}
  \label{tab:endmatter}
\end{center}
 \vspace{8mm}
\twocolumngrid
\end{enumerate}
\newpage

\begin{figure}[b!]
		\vspace{-2mm}
		\centering  \includegraphics[width=1\columnwidth]{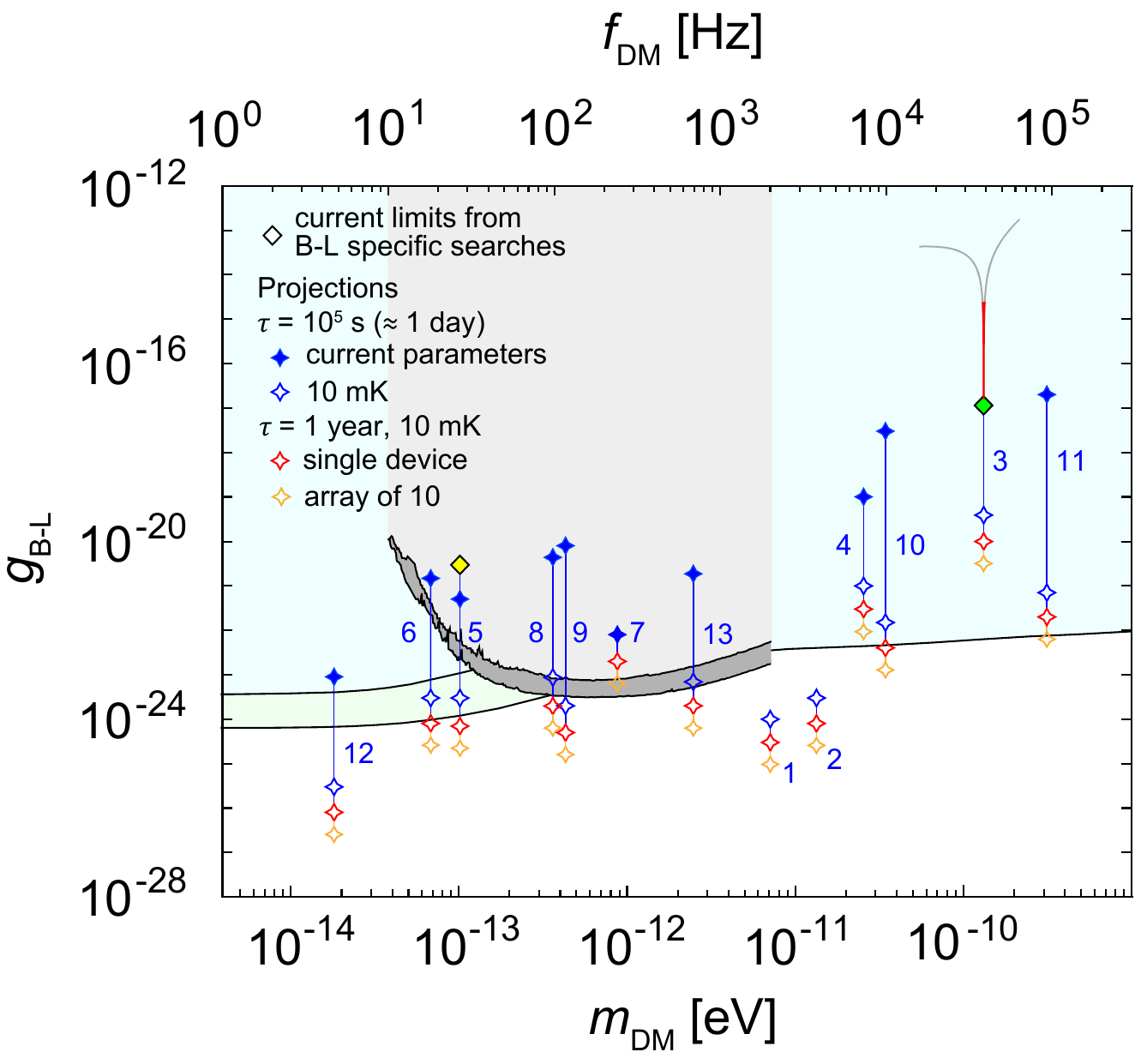}		\caption{Projected performance of contemporary optomechanical accelerometers as cryogenic UDM detectors, assuming an integration time of $\tau = 10^5$, a differential charge ratio of $\Delta_{12}=0.04$, and experimental parameters as described in Table 1.  Orange markers assume coherent averaging of 10 thermal-noise-limited sensors.} 
		\label{fig:5}
        \vspace{-2mm}
	\end{figure}

Projected thermal noise limited sensitivities $g_\t{B-L}^{\t{(th)}}$ assuming integration times of $\tau = 10^5\;\t{s}$ and $\tau = 1$ year, and a fractional charge ratio of $\Delta_{12} = 0.04$ are shown in Fig. \ref{fig:5}, overlaid on the experimental constraint plot in Fig. 4(d).  Here we use Eq. 2 (corresponding to a 1-sigma upper limit), which implicitly assumes the test mass has been cold-damped so that $\gamma \gtrsim \gamma_\t{DM}$.  We also include projections for an array of $N=10$ sensors, assuming $\sqrt{N}$ enhancement for coherent signal combination limited by (uncorrelated) thermal noise \cite{carney2021ultralight,brady2023entanglement}.
\newpage

\bibliography{ref}